\documentclass[conference]{IEEEtran}
\IEEEoverridecommandlockouts
\usepackage{cite}
\usepackage{amsmath,amssymb,amsfonts}
\usepackage{graphicx}
\usepackage{textcomp}
\usepackage{enumitem}
\usepackage{xcolor}
\usepackage{times}
\usepackage[binary-units=true]{siunitx}
\usepackage{latexsym}
\usepackage{hyperref}
\hypersetup{colorlinks=true,linkcolor=black,citecolor=blue,filecolor=black,urlcolor=blue}
\usepackage{algpseudocode}
\usepackage{algorithm}
\usepackage{caption}
\usepackage{subcaption}

\algnewcommand\algorithmicforeach{\textbf{for each}}
\algdef{S}[FOR]{ForEach}[1]{\algorithmicforeach\ #1\ \algorithmicdo}
\def\BibTeX{{\rm B\kern-.05em{\sc i\kern-.025em b}\kern-.08em
    T\kern-.1667em\lower.7ex\hbox{E}\kern-.125emX}}

\begin{document}

\title{A performance comparison of Dask and Apache Spark for data-intensive neuroimaging pipelines}

\author{Mathieu Dugr\'e, Val\'erie Hayot-Sasson, Tristan Glatard\\
Department of Computer Science and Software Engineering\\
Concordia University, Montr\'eal, Qu\'ebec, Canada\\
\{mathieu.dugre, valerie.hayot-sasson, tristan.glatard\}@concordia.ca
\vspace*{0.8cm} 
}

\maketitle

\begin{abstract}

In the past few years, neuroimaging has entered the Big Data era due to the joint
increase in image resolution, data sharing, and study sizes. However, no particular
Big Data engines have emerged in this field, and several alternatives remain
available. We compare two popular Big Data engines with Python APIs, Apache Spark and
Dask, for their runtime performance in processing neuroimaging pipelines. Our
evaluation uses two synthetic pipelines processing the \SI{81}{\giga\byte} BigBrain
image, and a real pipeline processing anatomical data from more than 1,000 subjects.
We benchmark these pipelines using various combinations of task durations, data
sizes, and numbers of workers, deployed on an 8-node (8 cores ea.) compute cluster in
Compute Canada's Arbutus cloud. We evaluate PySpark's RDD API against Dask's Bag,
Delayed and Futures. Results show that despite slight differences between Spark and
Dask, both engines perform comparably. However, Dask pipelines risk being limited
by Python's GIL depending on task type and cluster configuration. In all cases, the
major limiting factor was data transfer. While either engine is suitable for
neuroimaging pipelines, more effort needs to be placed in reducing data transfer time.
\end{abstract}


\section{Introduction}
The recent rise in data sharing and improved data collection strategies have brought
neuroimaging to the Big Data era~\cite{ALFAROALMAGRO:18, van2014human}. Existing
neuroimaging workflow engines, such as Nipype~\cite{Nipype:11}, are well suited for
processing the standard compute-intensive neuroimaging pipelines, but lack
incorporated Big Data strategies (i.e. in-memory computing, data locality, and
lazy-evaluation) to improve performance of the increasingly prevalent data-intensive
pipelines. As was noted in~\cite{hayot2019performance}, in-memory computing, coupled
with data locality, can bring significant performance improvements to data intensive
neuroimaging pipelines. In this work, we investigate how performance benchmarks
generalize across Big Data engines. In particular, we study the differences between
Dask~\cite{Dask:15} and Apache Spark~\cite{Spark:16}, for their suitability in the
processing of neuroimaging pipelines. Our goal is to test whether Spark or Dask has a
clear performance advantage to process Big neuroimaging Data. 

Spark and Dask both offer in-memory computing, data locality, and lazy evaluation,
which is common for Big Data engines. Both their schedulers operate dynamically,
which is good when task runtimes are not known ahead of time~\cite{Dask:15}. They
also provide rich, high-level programming APIs, and support a variety of
infrastructure schedulers, such as Mesos~\cite{hindman2011mesos}, YARN~\cite{vavilapalli2013apache}, or HPC clusters (Dask only). Over
these similarities, the engines have differences.

First and foremost, Spark is written in Scala while Dask is in Python.
Given the popularity of Python in scientific communities, this arguably
gives an edge to Dask due to data serialization costs from Python to Scala.
On the other hand, Python's Global Interpreter Lock (GIL) might reduce
parallelism in some cases. This difference in programming languages also
has qualitative implications. As part of the \href{http://scipy.org}{SciPy}
ecosystem, Dask provides almost transparent parallelization of applications
manipulating NumPy arrays or Pandas data frames. On the other hand, Spark's
Java, R and Python APIs allow to easily parallelize analyses that combine
these languages, with reduced performance loss. Our study focuses on
performance, although we recognize that other factors will also impact the
choice of an engine in practice.

Spark and Dask were both included in the evaluation reported
in~\cite{Mehta:17}, where a neuroimaging application
processed approximately \SI{100}{\giga\byte} of data. In this work, Dask was
reported to have a slight performance advantage over Spark. Overall, Dask's
end-to-end time (makespan) was measured to be up to 14\% faster than Spark,
due to ``more efficient pipelining'' 
and serialization time to
Python. 
Dask, however, was reported to have a larger startup
time than Spark. 
The analysis remained at a quite high level
though, leaving most of the observed performance difference unexplained. In
comparison, our study will provide a detailed analysis of performance
differences and similarities between Spark and Dask for neuroimaging data processing.

The next section details the design of our benchmarking experiments. We consider two
data-intensive neuroimaging applications: high-resolution imaging,
represented by the BigBrain data~\cite{Amunts:13}, and large functional
MRI studies, represented by data from the consortium for reliability and
reproducibility (CoRR~\cite{zuo2014open}). We test application pipelines
involving different patterns (map-only, map-reduce), and different types of
implementations (plain Python, command-line,
containerized). We evaluate performance on a dedicated cluster
representative of the ones used in today's data-intensive neuroimaging
studies, using the main Dask and Spark APIs. The other sections present
our results, discussion, and conclusions.

\section{Material and Methods}

\subsection{Engines}

\subsubsection{Apache Spark} Apache Spark is a widely-used 
general-purpose Big Data engine. Its main abstraction, the Resilient Distributed
Dataset (RDD)~\cite{RDD}, is a fault-tolerant, parallel collection of data elements.
It achieves fault tolerance through the recording of data lineage, the sequence of
operations used to modify the original data. The RDD is the basis of Spark's other
data structures, namely, Datasets and DataFrames. Datasets are similar to RDDs, but
additionally use Spark SQL's optimized execution engine to further improve
performance. DataFrames, used to process tabular data, are Datasets where the data is
organized into named-columns. While the DataFrame API exists in all the available
language APIs, Datasets are limited to Scala and Java.

As data transfers in Big Data workflows are an important source of performance
bottlenecks, Spark incorporates the concepts of data locality and in-memory
computing. Data locality, popularized by MapReduce~\cite{dean2008mapreduce},
schedules tasks as close as possible to where the data is stored. In-memory computing
ensures data is maintained in memory whenever possible, as writing large amounts of
data to slower storage devices may be costly. To reduce any unnecessary communication
and computation, Spark also included lazy evaluation, which builds the entire task
graph prior to execution to determine what needs to be computed.

Spark is compatible with three different schedulers: Spark
Standalone, YARN and Mesos.
The Spark Standalone scheduler is a simple default scheduler built into Spark.
In contrast, the YARN scheduler is primarily designed to schedule Hadoop-based 
workflows, whereas Mesos can be used to schedule
a variety of different workflows. As researchers would likely be
executing their workflows in HPC environments with neither YARN or Mesos
installed, we limit our focus to Spark's Standalone scheduler.

The Spark Standalone scheduler is composed of three main processes: the
\emph{master}, the \emph{workers} (\emph{slaves}, in Spark) and the
\emph{driver}. The \emph{master} coordinates resources provisioned by
\emph{worker} processes on the cluster. The application is submitted to the
\emph{driver} that in turn requests workers to the master and dispatches tasks
to them. A job is divided into stages to be executed in a different process onto
the workers. Each stage's operation is represented as a high-level task in the
computation graph. The Spark standalone scheduler uses a FIFO 
(First-In-First-Out) job scheduling policy, and it has
two execution modes: client mode, where the driver runs in a dedicated process
outside of the Spark cluster, and cluster mode, where the driver runs within
a worker process. Our experiments use client mode as cluster mode is not
available in PySpark.

Python is commonly selected as the programming language of choice in scientific
communities, and in particular, for our use case of neuroscience, where
numerous specialized Python libraries exist to study the data.
While serialization of Python to Java may lead to significant overheads, we
chose to focus on PySpark API due to its suitability for neuroimaging
research. 

We used Apache Spark v2.4.0.

\subsubsection{Dask} Dask is a Python-based Big Data engine that is becoming
increasingly popular in the scientific Python ecosystem. Like Spark, Dask avoids data transfers and needless computation and communication through
in-memory computing, data locality, and lazy evaluation. Dask workflows can 
further reduce data transfer costs by leveraging multithreading whenever 
communication is not bounded by Python's GIL. Dask relies on data 
lineage to achieve fault tolerance, however, unlike Spark, it does not require
operations to be coarse-grained. Furthermore, Dask is lightweight and modular,
allowing users to only install the components they require.

Dask has five main data structures:
\href{https://docs.dask.org/en/latest/array.html}{Array}, \href{https://docs.dask.org/en/latest/bag.html}{Bag},
\href{https://docs.dask.org/en/latest/dataframe.html}{DataFrame},
 \href{https://docs.dask.org/en/latest/delayed.html}{Delayed}, 
 and \href{https://docs.dask.org/en/latest/futures.html}{Futures}. A Dask
Array is used for the
processing of large arrays. It provides a distributed clone of the popular NumPy
library. Similar to Spark's RDD, a Dask
Bag is a parallel
collection of Python objects. It offers a programming abstraction similar to the
\href{https://toolz.readthedocs.io/en/latest/}{PyToolz library}. A Dask
Dataframe is a parallel
composition of
\href{http://pandas.pydata.org/pandas-docs/stable/reference/api/pandas.DataFrame.html}{Pandas Dataframes},
 used to process a large amount of tabular data.
Dask Delayed supports
arbitrary tasks
that do not fit in the Array, DataFrame or Bag APIs. Finally,
Dask Futures are similar to
Delayed in that they support arbitrary tasks, but they operate in real-time
rather than lazily. Except for Dask Bag, all APIs, by default, use
the local multithreaded scheduler. Dask Bag, instead, relies on the local
multiprocessing scheduler.  All Dask data structures, except for
Dask Array and Dask DataFrame, were used in our experiments.

The Dask graph is the internal representation of a Dask application to be
executed  by the scheduler. API operations generate multiple small tasks in
the computation graph, allowing an easier representation of complex
algorithms.

The Dask engine is compatible with numerous distributed schedulers, including 
\href{https://github.com/dask/dask-yarn}{YARN} and
\href{https://github.com/mrocklin/dask-mesos}{Mesos}, similarly to Spark. Dask 
also provides its own distributed scheduler, known as the Dask Distributed
scheduler. Although Dask is compatible with schedulers commonly found in HPC
clusters, we chose to use Dask's Distributed scheduler, to keep the environment
balanced between both engines.

In the Dask Distributed scheduler, a process called \textit{dask-scheduler}
administrates the resources provided by \textit{workers} in the cluster. The
scheduler receives jobs from clients and assigns tasks to available workers.
Similarly to Spark's scheduler, Dask Distributed completes the processing of a
graph branch before moving along to the next one. 

In our experiments, we used Dask v1.1.4.

\subsection{Infrastructure}

 We used Compute Canada's
 \href{https://docs.computecanada.ca/wiki/Cloud\_resources}{Arbutus Cloud} operated by
 the \href{https://www.westgrid.ca}{WestGrid} regional organization at the University
 of Victoria, running OpenStack Queens release. We used c8-30gb-186 cloud
 instances with 8 VCPUs, an Intel Xeon Gold 6130 processor, \SI{30}{\giga\byte} of
 RAM at \SI{2666}{\mega\hertz}, \SI{20}{\giga\byte} of mounted storage, and a base
 image running CentOS 7.5.1804 with Linux kernel version
 3.10.0\-862.11.6.el7.x86\_64. Instances were connected by a
 \SI{10}{\giga\bit/\second} Ethernet network.
 
 Cloud instances hosted a single Dask or Spark worker, configured to use 8 CPUs. We
 used Dask's default configuration that uses all the available memory on the
 instance. Its default heuristic is to: target a 60\% memory load, spill to disk at
 70\%, pause the worker at 80\%, and terminate the worker at 95\%. We configured
 Spark to use 1 executor per worker and \SI{25}{\giga\byte} of memory per executor, to leave \SI{5}{\giga\byte}
 for off-heap. We configured the Spark driver to use \SI{25}{\giga\byte} of memory, and used the
 default configuration for the master. We used the default configuration for worker
 memory management: at 60\% it spills data to disk, and 50\% of that amount is
 reserved for storage that is immune to eviction.
 
 One cloud instance did not host any worker and had a \SI{2}{\tera\byte} disk volume
 shared with the other instances using the Network File System (NFS) v4.
 This instance was also used for the Spark driver and master, for the
 Dask scheduler, and for job monitoring with the Spark and Dask user
 interfaces. For both Spark and Dask, spilled data was evicted to the NFS.

\subsection{Dataset}

We used BigBrain~\cite{Amunts:13}, a three-dimensional image of a human
brain with voxel intensities ranging from 0 to 65,535. The original data is
stored in 125 blocks in the MINC~\cite{minc} HDF5-based format, available
at \url{ftp://bigbrain.loris.ca/BigBrainRelease.2015/3D\_Blocks/40um} at a
resolution of \SI{40}{\micro\metre}. We converted the blocks into the
\href{https://nifti.nimh.nih.gov/nifti-1}{NIfTI} format, a popular format
in neuroimaging. We left the NIfTI blocks uncompressed, resulting in 
a total data size of \SI{81}{\giga\byte}. 
To evaluate the effect of block size, we resplit these blocks into 30, 125 and 750 blocks of 
\SI{2.7}{\giga\byte}, \SI{0.648}{\giga\byte}, and
\SI{0.108}{\giga\byte}, using the \href{https://github.com/big-data-lab-team/sam}{sam} library.

We also used the dataset provided by the Consortium for Reliability and
Reproducibility (\href{http://fcon_1000.projects.nitrc.org/indi/CoRR/html/}{CoRR}~\cite{zuo2014open}) as
available on \href{http://datasets.datalad.org/?dir=/corr/RawDataBIDS}{DataLad}. The
entire dataset is \SI{408.4}{\giga\byte}, containing anatomical, diffusion and
functional images of 1,397 subjects acquired in 29 sites. We used all 3,491 anatomical
images, representing \SI{39}{\giga\byte} overall (\SI{11.17}{\mega\byte} per image on
average).

\subsection{Applications}

We used three neuroimaging applications to evaluate the engines in different
conditions. The first two, incrementation and histogram, are simple 
synthetic applications representing map-only and map-reduce applications,
respectively. The third, is a real neuroimaging application representative
of popular BIDS applications~\cite{gorgolewski2017bids}. All scripts used for our
experiments are available at
\url{https://github.com/big-data-lab-team/paper-big-data-engines}

\subsubsection{Incrementation}
\begin{figure}[!b]
    \centering
    \includegraphics[height=\columnwidth,
    angle=-90]{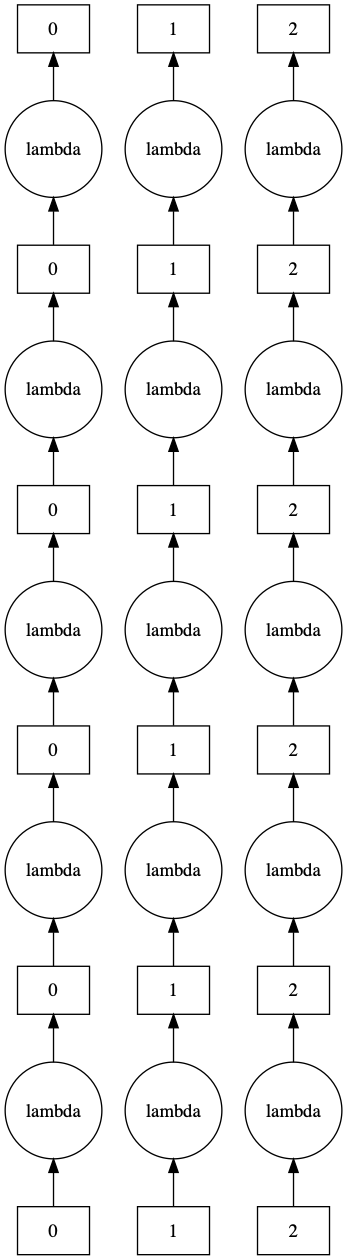}
    \caption{Task graph for Incrementation with 5 iterations and 3 BigBrain blocks.
    Circles represent the incrementation and sleep function while rectangles
    represent data elements.}\label{fig:tg-inc}
\end{figure}

We used an adaptation of the image incrementation pipeline used
in~\cite{hayot2019performance} (see Algorithm~\ref{alg:incrementation}).
The application reads blocks of the BigBrain image from the shared file
system, increments the intensity value of each voxel by 1 to avoid caching
effects, sleeps for a configurable amount of time to emulate more
compute-intensive processing, repeats this process for a specified number of
iterations, and
finally writes the result as a NIfTI image back to the shared file system.
This application allows us to study the behavior of the engines when all
inputs are processed independently, in a map-only scenario (see
Figure~\ref{fig:tg-inc}). This mimics the behavior of analyzing multiple
independent subjects in parallel.

\begin{algorithm}[!b]
    \caption{Incrementation (adapted from~\cite{hayot2019performance})}\label{alg:incrementation}
    \begin{algorithmic}
    \Require{\(x\), a sleep delay in float}
    \Require{\(file\), a file containing a BigBrain block}
    \Require{\(fs\), NFS mount to write image to}
    \State{Read \(block\) from \(file\)}
    \ForEach{\(i \in iterations\)}
        \ForEach{\(block \in image\)}
            \State{\(block\gets block+1\)}
            \State{Sleep \(x\)}
        \EndFor
    \EndFor
    \State{Write \(block\) to \(fs\)}
\end{algorithmic}
\end{algorithm}

\subsubsection{Histogram}

\begin{figure}[!t]
    \centering
    \includegraphics[height=\columnwidth,
    angle=-90]{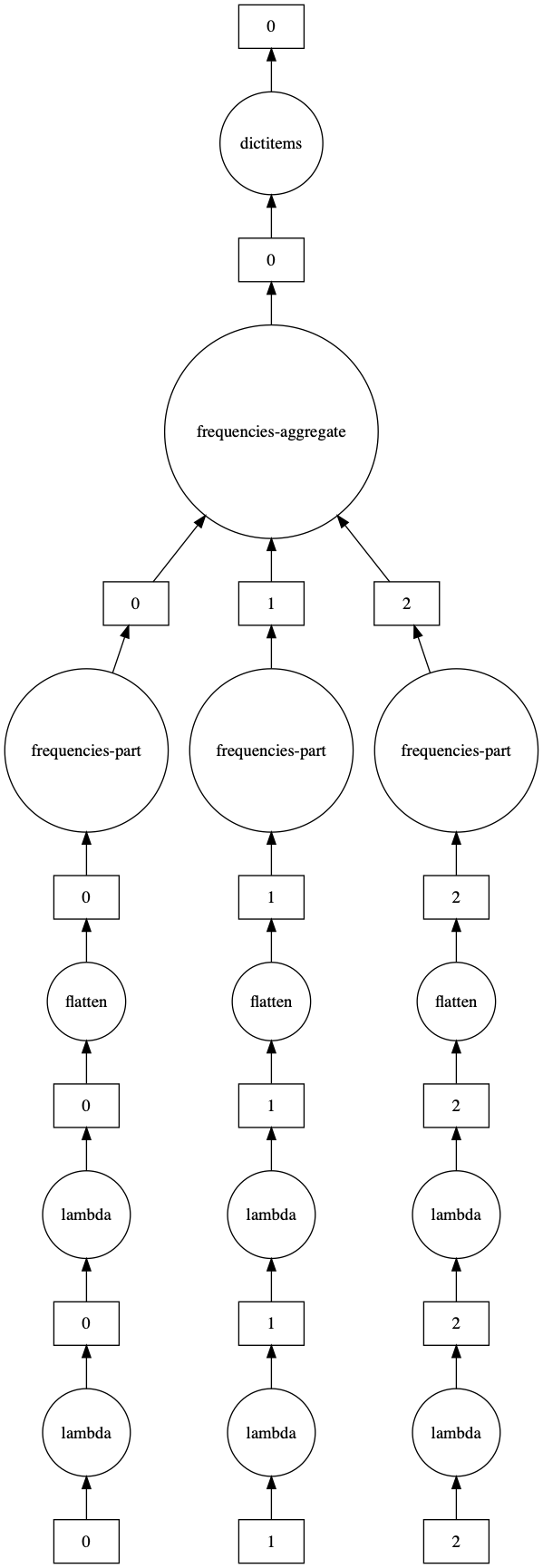}
    \caption{Task graph for Histogram with 3 BigBrain blocks. Circles represent the
    functions while rectangles represent data elements.
    }\label{fig:tg-histo}
\end{figure}

 Our second application calculates the histogram of the BigBrain image (see
 Algorithm~\ref{alg:histogram}). It reads the image blocks from the shared
 file system, calculates intensity frequencies, aggregates the frequencies
 across the blocks, and finally writes the resulting histogram to the
 shared file system as a single \SI{766}{\kilo\byte} file. It follows the
 map-reduce paradigm, in which the final result is obtained from all the
 input blocks (see Fig.~\ref{fig:tg-histo}). This application requires data
 shuffling, incurring additional performance costs compared to
 Algorithm~\ref{alg:incrementation}. The total amount of shuffled data is,
 however, limited to \SI{2.62}{\mega\byte} per block as it only consists
 of image histograms. Two implementations are studied: a pure Python one,
 where frequencies are computed through Python dictionary manipulations, and one based
 on the \href{https://github.com/numpy}{NumPy} library, that implements array computations in C.

\begin{algorithm}[!t]
    \caption{Histogram}\label{alg:histogram}
    \begin{algorithmic}
    \Require{\(files\), files containing BigBrain blocks}
    \Require{\(fs\), NFS mount to save image to}
    \ForEach{\(file \in files\)}
        \State{Read \(block\) from \(file\)}
        \State{{Calculate \(frequencies\) in \(block\)}}
    \EndFor
    
    \State{\(histogram\gets\)Aggregate \(frequencies\) of each \(file\)}

    \State{Write \(histogram\) to \(fs\)}
    \end{algorithmic}
\end{algorithm}

\subsubsection{BIDS app example}

We used the
\href{https://github.com/BIDS-Apps/example}{BIDS app example} map-reduce style 
application. The map phase, also called participant analysis, extracts the brain
tissues of a subject's 3D MRI from the CoRR dataset using the FMRIB Software Library
(\href{https://fsl.fmrib.ox.ac.uk/fsl/fslwiki}{FSL}), and writes the resulting
image (\SI{2.47}{\mega\byte} on average) to the NFS. The reduce phase, or group
analysis, computes the volume of each brain and returns the average
volume, shuffling a total of \SI{8.6}{\giga\byte} image data.

While Incrementation and Histogram were implemented directly in Python, the 
BIDS App example requires an external set of command-line tools distributed as a Docker container image
(bids/base\_fsl on DockerHub). We converted the Docker image to a Singularity image for use in 
an HPC environment, using
\href{https://hub.docker.com/r/singularityware/docker2singularity/tags/}{docker2singularity}.
The image was preloaded on the NFS and Singularity version 3.2.1\-1.el7 was installed
for all instances.

\subsection{Experiments}

Four parameters were varied in our experiments, as shown in
Table~\ref{tab:param}. We varied (1) the number of workers, to assess the
scalability of the engines, (2) the number of BigBrain blocks in
Incrementation and Histogram, to measure the effect of different I/O patterns
and parallelization degrees, and (3) the number of iterations and sleep
delay in Incrementation, to evaluate the effect of job length and number of
tasks.
It should be noted that increasing the number of blocks or iterations also
increases the total compute time of the application for a given sleep
delay. To avoid any potential external bias such as background load on the
network, we ran the experiments in a randomized order and cleared the page
cache of each worker before each execution.

For each run, we measured the application makespan as well as the cumulative 
data read, compute, data write, and engine overhead times across all application
tasks. 
The overhead calculation for each CPU is the end time of its last processed task
minus the total runtime of the tasks it executed. The summation of those results is the
total overhead.


\begin{table}[!t]
    \renewcommand{\arraystretch}{1.3}
    \caption{Parameters for the experiments}\label{tab:param}
    \centering
    \begin{tabular*}{\columnwidth}{llll}
    \hline
                        & Incrementation & Histogram             & BIDS Example          \\ \hline
    \# of worker        & 1, 2, 4, 8     & 1, 2, 4, 8            & 1, 2, 4, 8            \\
    \# of blocks        & 30, 125, 750   & 30, 125, 750          & \multicolumn{1}{c}{-} \\
    \# of iterations    & 1, 10, 100     & \multicolumn{1}{c}{-} & \multicolumn{1}{c}{-} \\
    Sleep delay {[}\SI{}{\second}{]} & 1, 4, 16, 64   & \multicolumn{1}{c}{-} & \multicolumn{1}{c}{-} \\ \hline
    \end{tabular*}
    \end{table}

\section{Results}

\subsection{Incrementation: Number of workers}
\begin{figure}[!b]
    \centering
    \begin{subfigure}[b]{\columnwidth}
        \includegraphics[clip,width=\columnwidth]{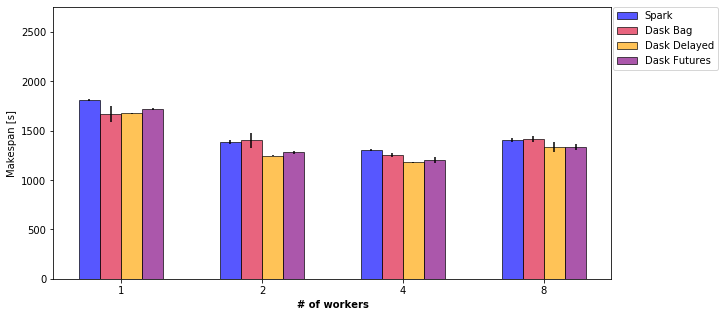}%
        \caption{Incrementation makespan}\label{fig:inc_ms_worker}
    \end{subfigure}
    \\
    \begin{subfigure}[b]{\columnwidth}
        \includegraphics[clip,width=\columnwidth]{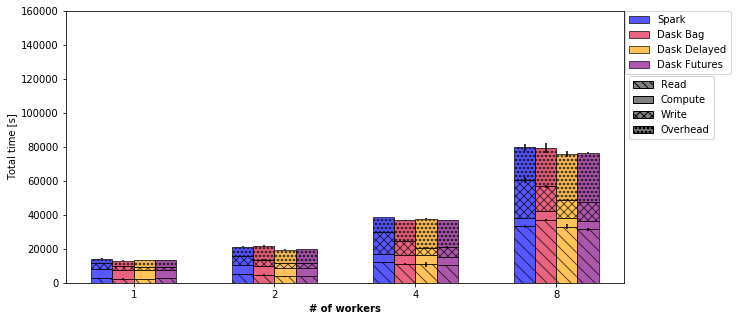}%
        \caption{Incrementation total time}\label{fig:inc_tt_worker}
    \end{subfigure}
    \caption{125 blocks, 10 iterations, \SI{4}{\second} delay, 8
    CPUs/worker}\label{fig:inc_worker}
\end{figure}

Figure~\ref{fig:inc_ms_worker} shows the makespan of the Incrementation application
for different numbers of workers and engines. The bars show the average
makespan over 3 repetitions while the error bars are the standard deviations. Overall,
there is no substantial makespan difference between the engines. Dask seems to have a
slight advantage over Spark, \SI{83.61}{\second} on average,
with Delayed and Futures being slightly better than Bag.

For all engines, the makespan is far from decreasing linearly with the
number of workers. The makespan
even increases between 4 and 8 workers. This is due to the high impact of
data transfers and engine overhead on the application. 

Figure~\ref{fig:inc_tt_worker} shows the total execution time of the
Incrementation application, broken down into data transfer (read and
write), compute (sleep), and overhead time. As expected, the computing time
stays similar when the number of workers increases. However, the data
transfer time and overhead increase proportionally to the number of
workers with regression slopes: Spark (\SI{2337}{\second/task}), Bag
(\SI{2650}{\second/task}), Delayed (\SI{3251}{\second/task}), Futures
(\SI{3570}{\second/task}).

On Figure~\ref{fig:inc_tt_worker}, we also note that Spark's overhead is
slightly lower than Dask's, particularly when the number of workers
increases. Moreover, Delayed and Futures have a higher overhead than Bag.
However, overhead differences are compensated by an increase in data
transfer time, as a reduced overhead increases the concurrency between data
transfers and the application saturates the disk bandwidth of the shared file system.

\subsection{Incrementation: Number of blocks}

\begin{figure}[!t]
    \centering
    \begin{subfigure}[b]{\columnwidth}
        \includegraphics[clip,width=\columnwidth]{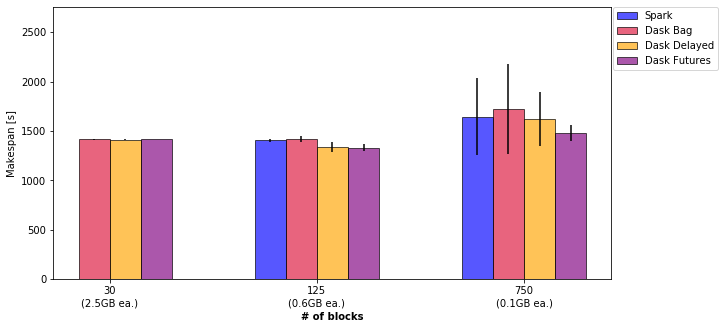}%
        \caption{Incrementation makespan}\label{fig:inc_ms_block}
    \end{subfigure}
    \\
    \begin{subfigure}[b]{\columnwidth}
        \includegraphics[clip,width=\columnwidth]{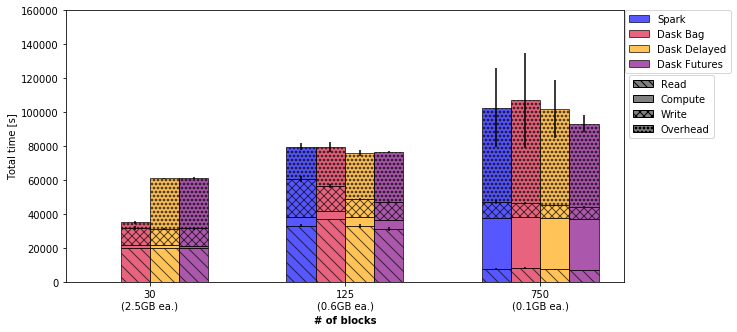}%
        \caption{Incrementation total time}\label{fig:inc_tt_block}
    \end{subfigure}
    \caption{10 iterations, \SI{4}{\second} delay, 8 workers, 8
    CPUs/worker}\label{fig:inc_block}
\end{figure}

Figure~\ref{fig:inc_ms_block} shows the Incrementation makespan when varying the
number of image blocks for constant BigBrain image size. We were not able to run
Spark for 30 blocks due to its \SI{2}{\giga\byte} limitation in the task size.
 Once again, we do not observe any substantial difference among the engines.
For all engines, makespan variability increases with the number of blocks, however,
engines scale very well in general.

In Figure~\ref{fig:inc_tt_block}, the total execution time of each function is shown.
For 30 blocks, the increased overhead of Delayed and Futures is a calculation artifact 
coming from the fact that only 30 of the 64 available threads can be used concurrently. Once again, the data transfer time reduces with more blocks but 
the overhead time increases by a similar amount. This is not observed for 30 blocks
as the workers are not used at full capacity, i.e., some threads are idle. Finally,
the variability of the overhead increases with the number of blocks, which explains
the makespan variability mentioned previously.

\subsection{Incrementation: Number of iterations}
\begin{figure}[!t]
    \centering
    \begin{subfigure}[b]{\columnwidth}
        \includegraphics[clip,width=\columnwidth]{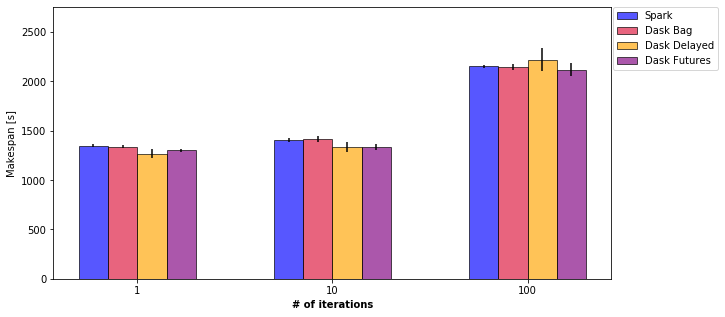}%
        \caption{Incrementation makespan}\label{fig:inc_ms_itr}
    \end{subfigure}
    \\
    \begin{subfigure}[b]{\columnwidth}
        \includegraphics[clip,width=\columnwidth]{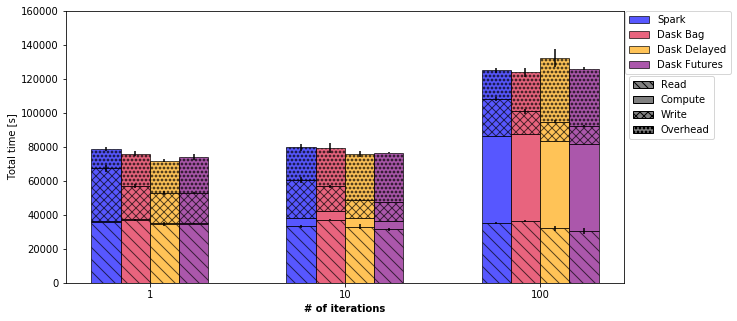}%
        \caption{Incrementation total time}\label{fig:inc_tt_itr}
    \end{subfigure}
    \caption{125 blocks, \SI{4}{\second} delay, 8 workers, 8 CPUs/worker}
\end{figure}

Figure~\ref{fig:inc_ms_itr} shows the makespan of the application while
varying the number of iterations. Overall, Spark and Dask APIs are once
again equivalent, although Delayed and Futures are slightly faster
than Bag and Spark for 1 and 10 iterations, and Futures are faster than
Delayed, Bag and Spark for 100 iterations. Differences remain minor
though.

In Figure~\ref{fig:inc_tt_itr}, the total execution time breakdown is
shown. We observe the good scalability of all the engines with the number of
iterations.

\subsection{Incrementation: Sleep delay}
\begin{figure}[!b]
    \centering
    \begin{subfigure}[b]{\columnwidth}
        \includegraphics[clip,width=\columnwidth]{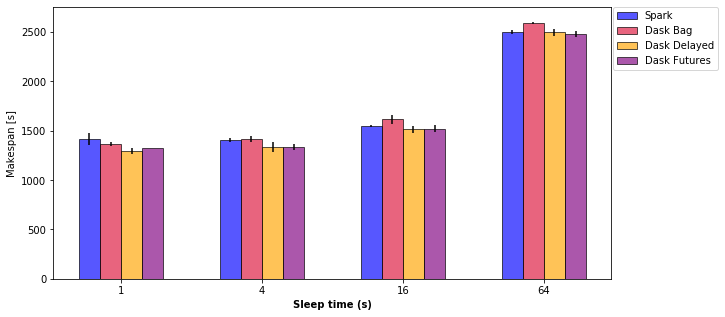}%
        \caption{Incrementation makespan}\label{fig:inc_ms_sleep}
    \end{subfigure}
    \\
    \begin{subfigure}[b]{\columnwidth}
        \includegraphics[clip,width=\columnwidth]{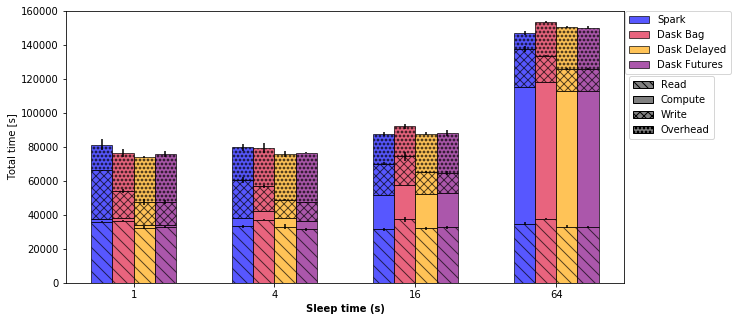}%
        \caption{Incrementation total time}\label{fig:inc_tt_sleep}
    \end{subfigure}
    \caption{125 blocks, 10 iterations, 8 workers, 8 CPUs/worker}
\end{figure}

Figure~\ref{fig:inc_ms_sleep} shows the makespan of the Incrementation application
for different sleep delays. Overall, all engines again perform the same and scale
well with task duration. Spark is initially slower than the Dask APIs, however, it is
faster with an increased sleep delay. Also, within the Dask APIs, Dask Bag is
slower than the other two, but it is not considerable.

Figure~\ref{fig:inc_tt_sleep} shows the total execution time breakdown. Spark has the
smallest overhead. As previously observed, variations in overhead time are almost
exactly compensated by variations in data transfer time.

\subsection{Incrementation: Gantt chart}

\begin{figure}[!b]
    \centering
    \begin{subfigure}[b]{\columnwidth}
        \href{https://mathdugre.github.io/paper-big-data-engines/spark-inc-baseline.html}{
            \includegraphics[clip,width=\columnwidth,
            height=0.15\textheight]{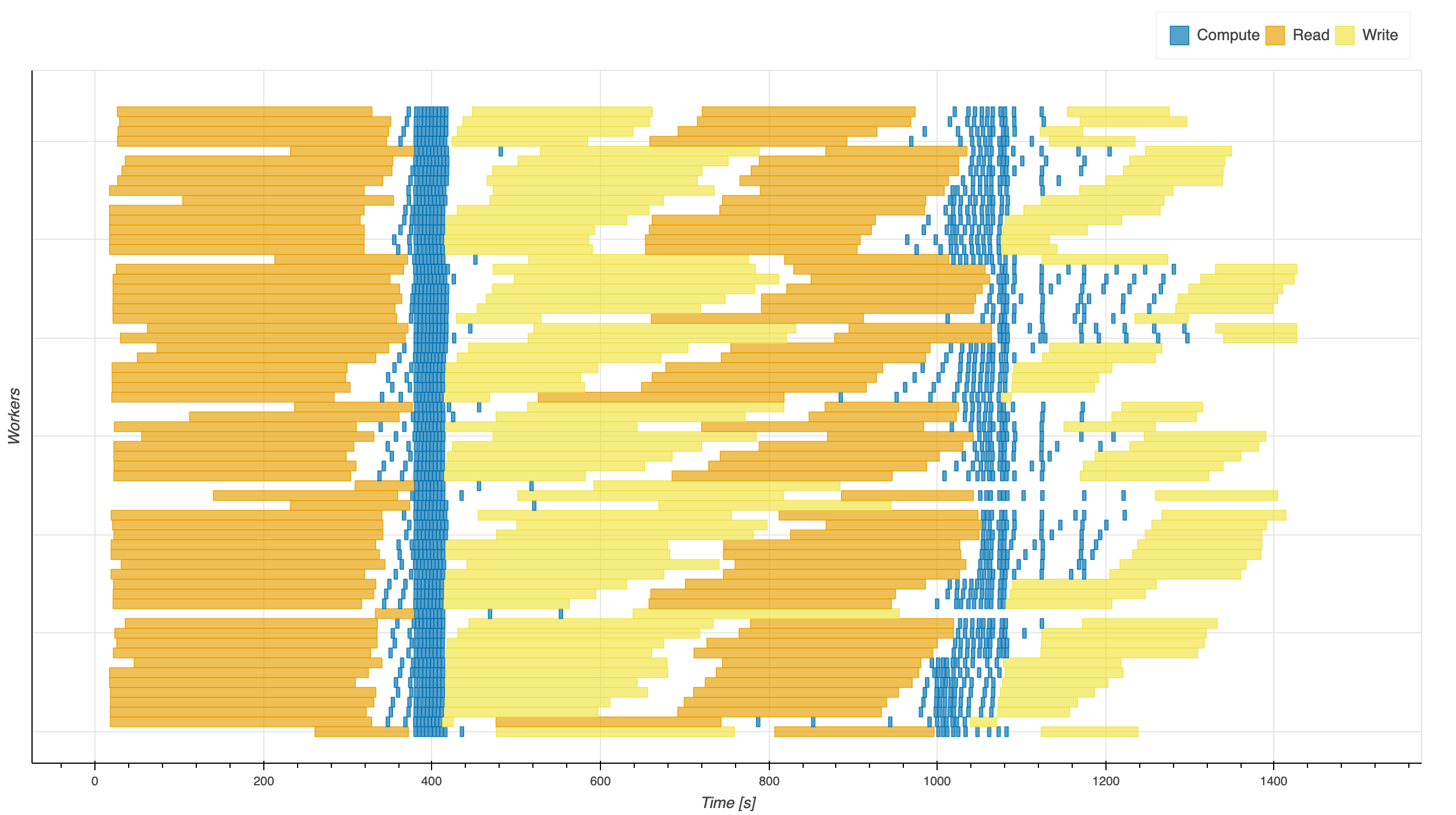}}
        \caption{Spark}\label{fig:inc_spark_gantt}
    \end{subfigure}
    \\
    \begin{subfigure}[b]{\columnwidth}
        \href{https://mathdugre.github.io/paper-big-data-engines/dask-bag-inc-baseline.html}{
        \includegraphics[clip,width=\columnwidth,
        height=0.15\textheight]{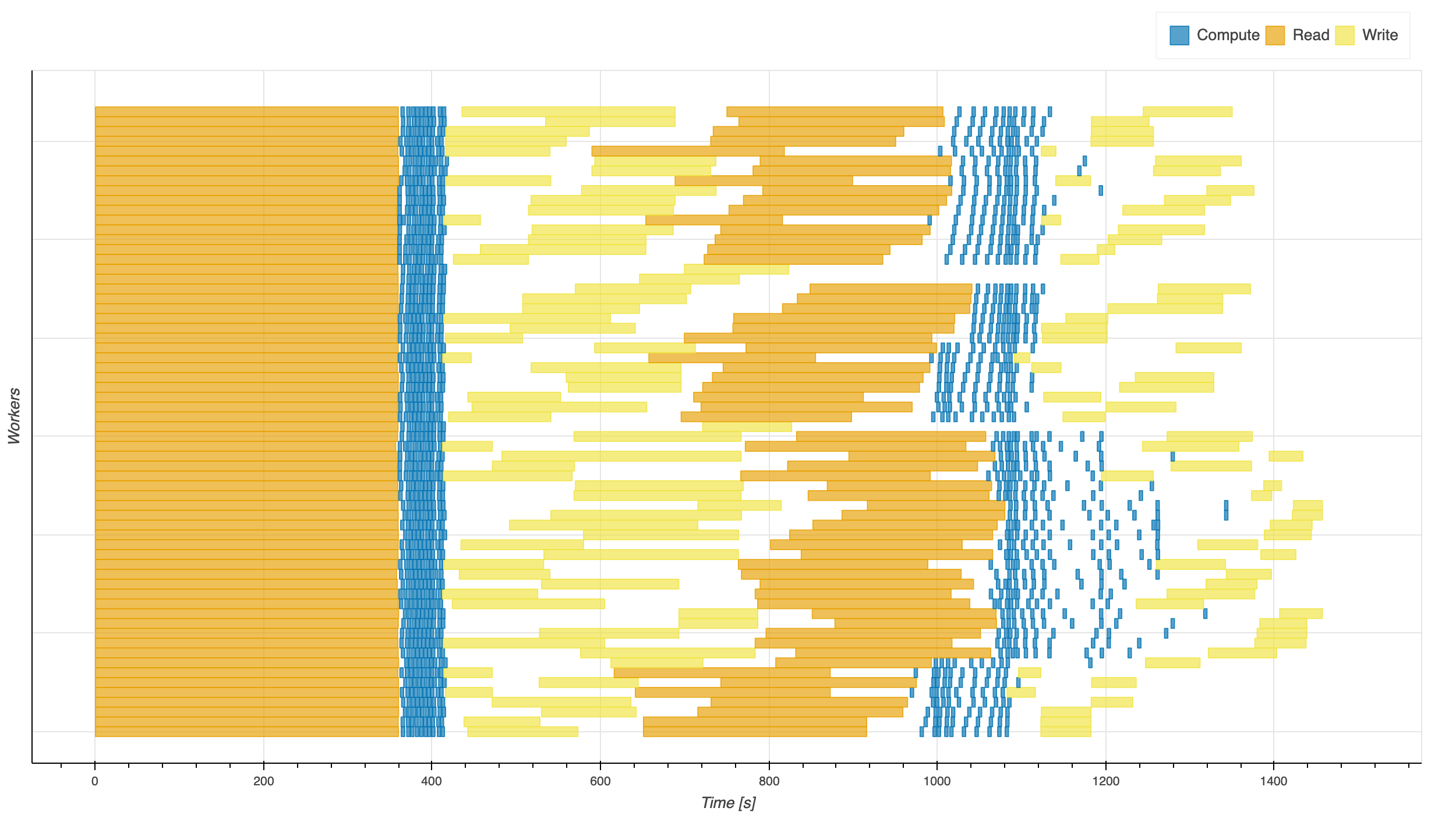}}
        \caption{Dask Bag}\label{fig:inc_dask_bag_gantt}
    \end{subfigure}
    \\
    \begin{subfigure}[b]{\columnwidth}
        \href{https://mathdugre.github.io/paper-big-data-engines/dask-delayed-inc-baseline.html}{
        \includegraphics[clip,width=\columnwidth,
        height=0.15\textheight]{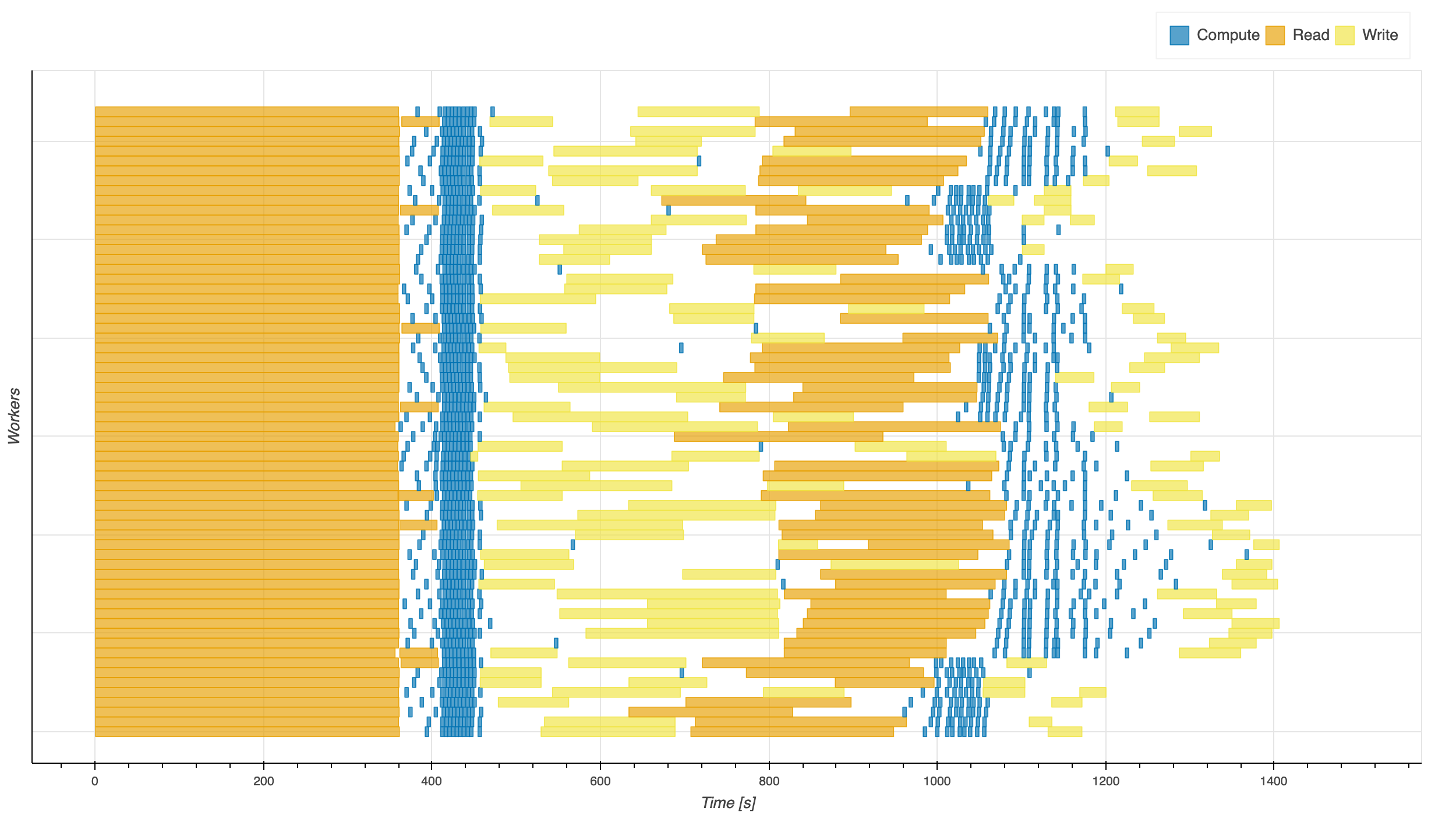}}
        \caption{Dask Delayed}\label{fig:inc_dask_delayed_gantt}
    \end{subfigure}
    \\
    \begin{subfigure}[b]{\columnwidth}
        \href{https://mathdugre.github.io/paper-big-data-engines/dask-futures-inc-baseline.html}{
        \includegraphics[clip,width=\columnwidth,
        height=0.15\textheight]{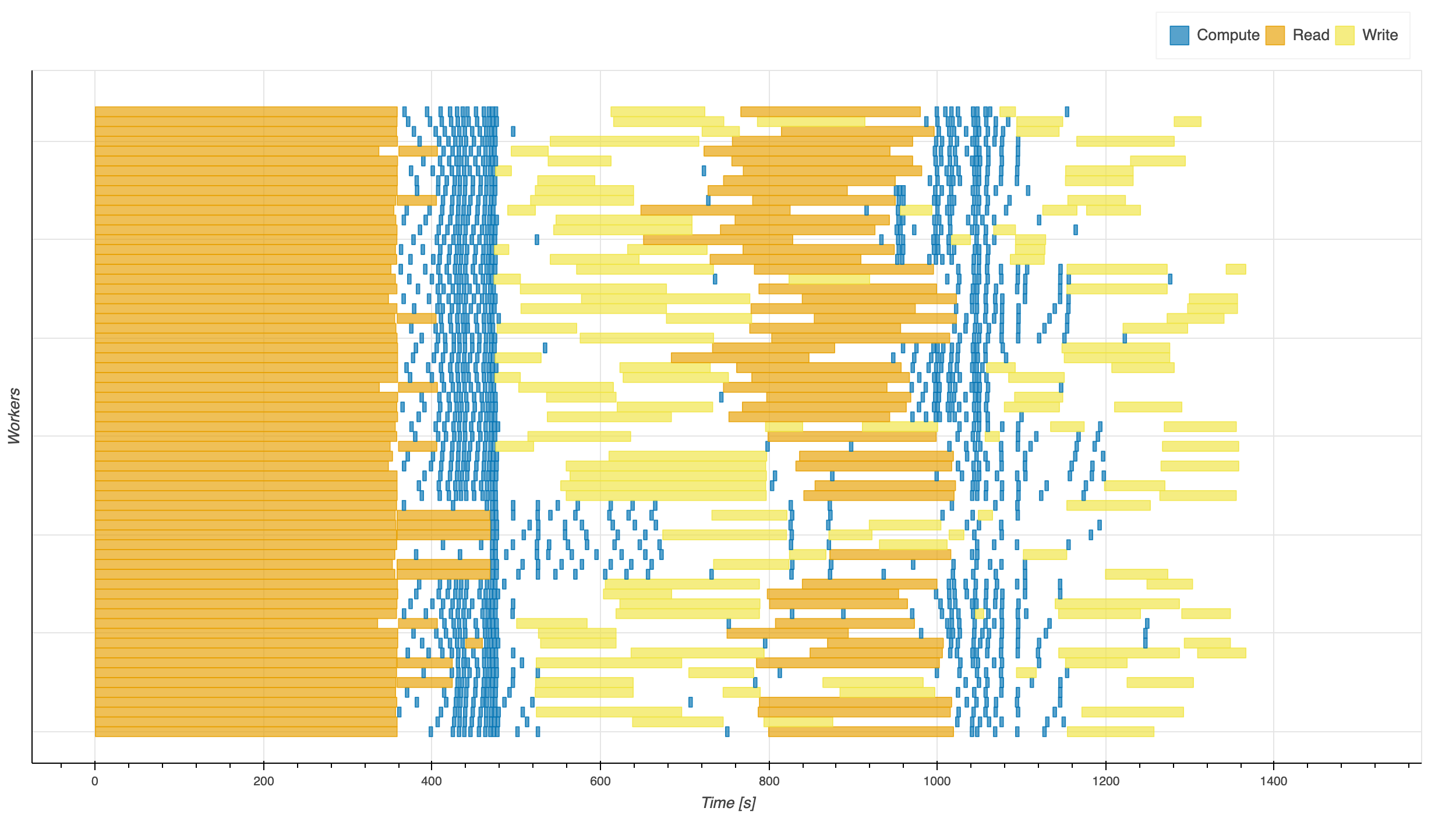}}
        \caption{Dask Futures}\label{fig:inc_dask_futures_gantt}
    \end{subfigure}
    \caption{Incrementation Gantt chart: 125 blocks, 10 iterations, \SI{4}{\second} delay, 8 workers,
8 CPUs/worker. All Gantt charts are clickable and link to interactive figures with additional information.}\label{fig:inc_gantt}
\end{figure}

Figure~\ref{fig:inc_gantt} shows the Gantt chart obtained for each engine. Gantt
charts are structured in batches of up to 64 read-compute-write concurrent sequences.
File reads in the first batch are much longer than the ones in the following batches:
this is due to the high synchronization of data transfers that leads to a high
saturation of the shared file system. We also note that overhead, represented in
white, is concentrated around the data transfer tasks and the
computing tasks that run concurrently with data transfers. 

\subsection{Histogram (Python): Number of workers}
\begin{figure}[!b]
    \centering
    \begin{subfigure}[b]{\columnwidth}
        \includegraphics[clip,width=\columnwidth]{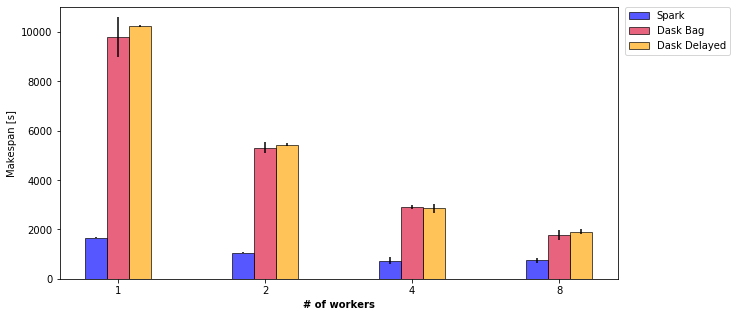}%
        \caption{Histogram (Python) makespan}\label{fig:histo_ms_worker}
    \end{subfigure}
    \\
    \begin{subfigure}[b]{\columnwidth}
        \includegraphics[clip,width=\columnwidth]{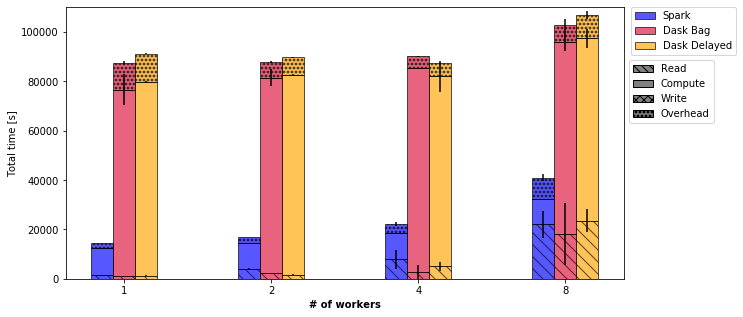}%
        \caption{Histogram (Python) total time}\label{fig:histo_tt_worker}
    \end{subfigure}
    \caption{125 blocks, 8 CPUs/worker}\label{fig:histo_worker}
\end{figure}

Figure~\ref{fig:histo_ms_worker} shows the makespan of the Histogram application for various
amounts of workers. Spark is significantly faster than Dask APIs. The difference
narrows as the number of workers increases but it remains significant. Between Dask APIs
there is no substantial difference, however, Bag is slightly faster on average.

From Figure~\ref{fig:histo_tt_worker}, the total time spent in each function is
shown. The computing time is significantly larger for Dask than for Spark. This is
presumably due to Python's GIL preventing Dask to parallelize the computation
over
multiple threads. Overall, the I/O and overhead times are comparable for all engines.

Figure~\ref{fig:histo_tt_worker} shows that Dask engines have a similar
total execution time when their number of workers varies, except for 8
workers where it increases slightly due to increased read time.
On the other hand, Spark total execution time keeps increasing; especially at 8
workers. This is because Dask engines benefit more from additional workers, in
this GIL-bounded scenario.

\subsection{Histogram (Python): Number of blocks}
\begin{figure}[!b]
    \centering
    \begin{subfigure}[b]{\columnwidth}
        \includegraphics[clip,width=\columnwidth]{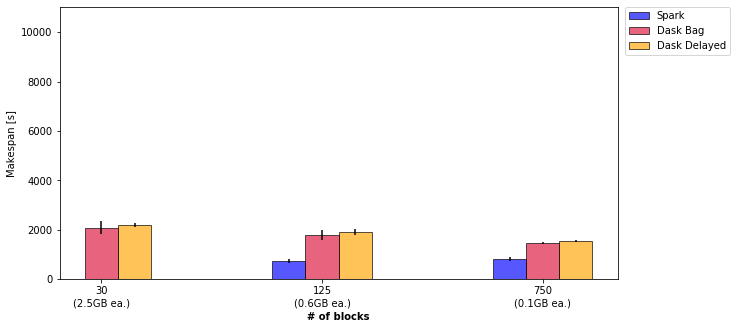}%
        \caption{Histogram (Python) makespan}\label{fig:histo_ms_block}
    \end{subfigure}
    \\
    \begin{subfigure}[b]{\columnwidth}
        \includegraphics[clip,width=\columnwidth]{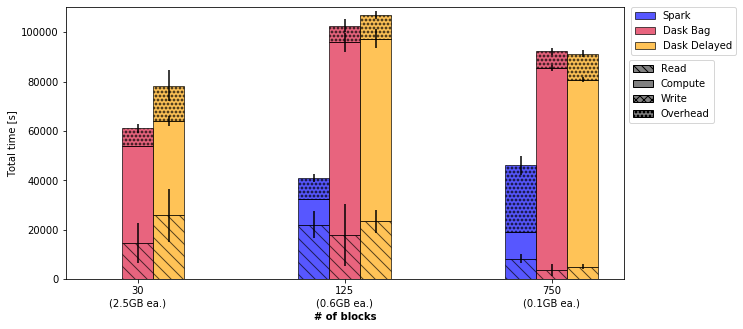}%
        \caption{Histogram (Python) total time}\label{fig:histo_tt_block}
    \end{subfigure}
    \caption{8 workers, 8 CPUs/worker}\label{fig:histo_block}
\end{figure}

Figure~\ref{fig:histo_ms_block} shows the makespan for different block sizes. 
Once more, Spark is significantly faster than Dask engines and Bag is slightly
faster than Delayed. Overall, the engines do not react to changes in block size.

In Figure~\ref{fig:histo_tt_block} the total time for each function is shown.
Note that at 30 blocks the workers' resources are not fully used since the
workers can process up to 64 tasks in parallel. Moreover, the overhead for
30 blocks is erroneous due to a lower amount of block than available threads.
This leads to thread based engines to have higher overhead. Given that, 
Spark has a much lower compute time which makes it significantly faster
than Dask engines. Overall, for all engines, lower block size results in lower
data transfer time but increases the overhead time. Again, this is due to large overhead
desynchronizing task thus reducing the NFS bottleneck.

\subsection{Histogram (Python): Gantt chart}
\begin{figure}[!b]
    \centering
    \begin{subfigure}[b]{\columnwidth}
        \href{https://mathdugre.github.io/paper-big-data-engines/spark-histo-baseline.html}{
        \includegraphics[clip,width=\columnwidth,
        height=0.15\textheight]{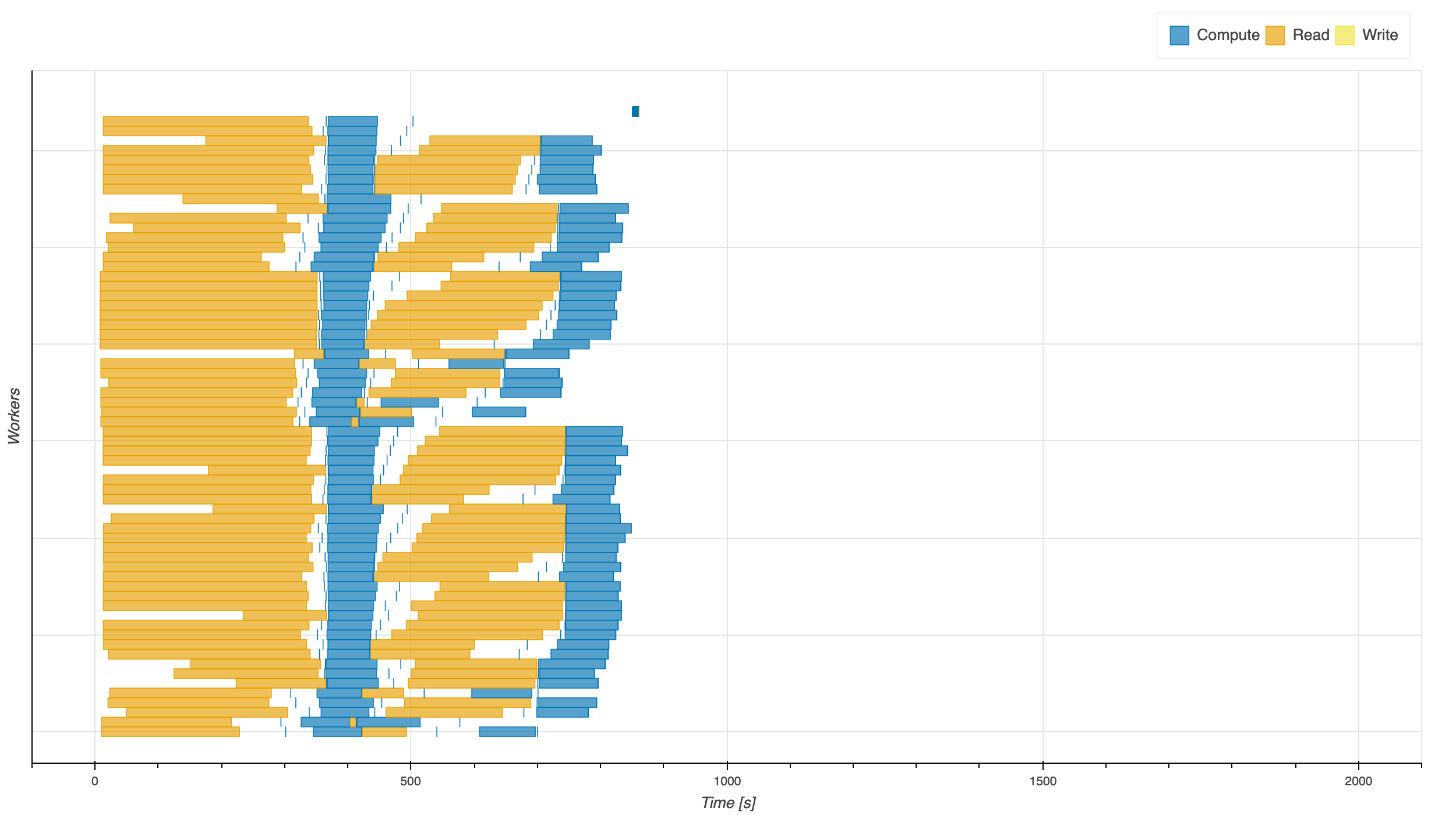}}
        \caption{Spark}\label{fig:histo_spark_gantt}
    \end{subfigure}
    \\
    \begin{subfigure}[b]{\columnwidth}
        \href{https://mathdugre.github.io/paper-big-data-engines/dask-bag-histo-baseline.html}{
        \includegraphics[clip,width=\columnwidth,
        height=0.15\textheight]{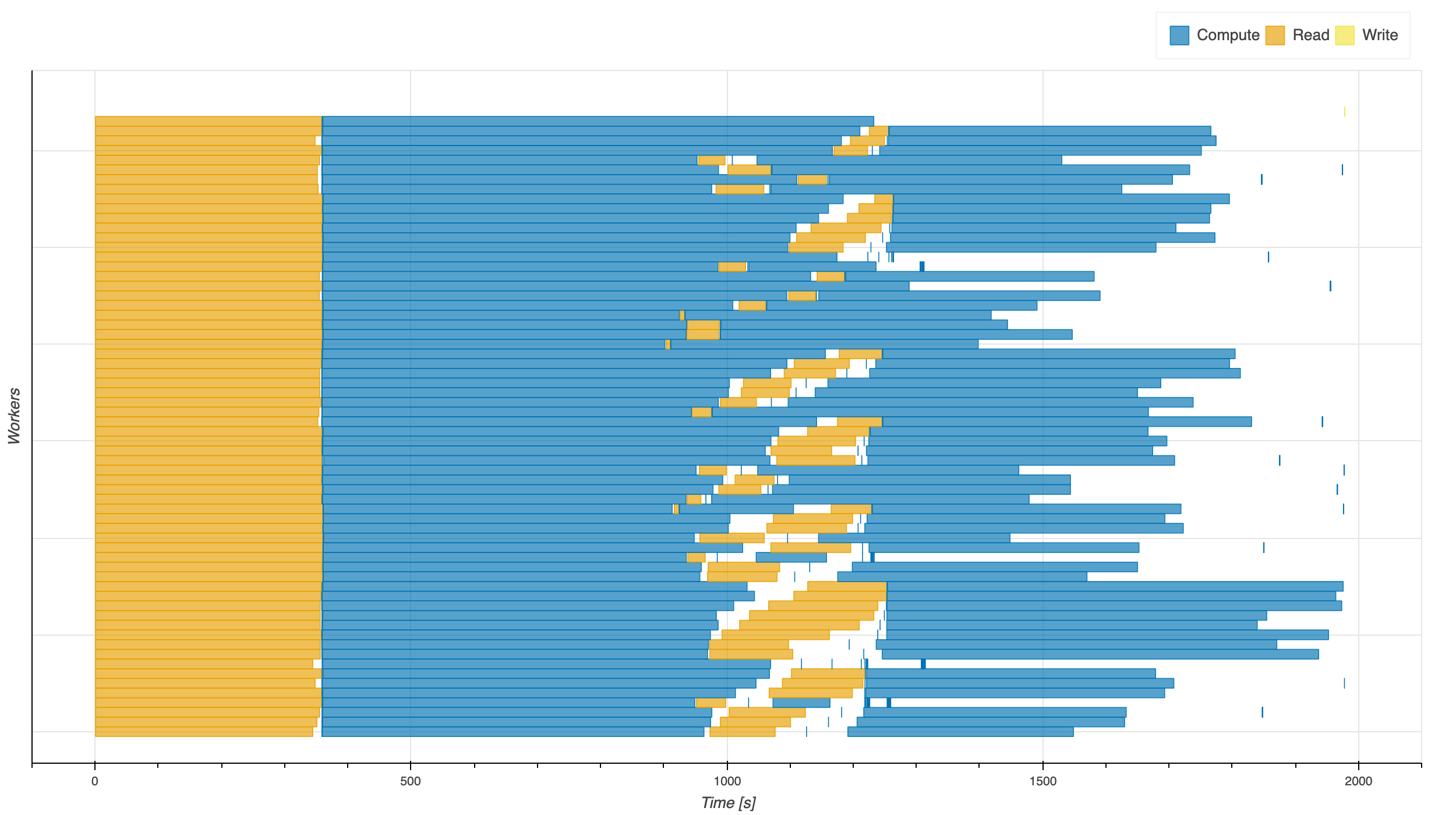}}
        \caption{Dask Bag}\label{fig:histo_dask_bag_gantt}
    \end{subfigure}
    \\
    \begin{subfigure}[b]{\columnwidth}
        \href{https://mathdugre.github.io/paper-big-data-engines/dask-delayed-histo-baseline.html}{
        \includegraphics[clip,width=\columnwidth,
        height=0.15\textheight]{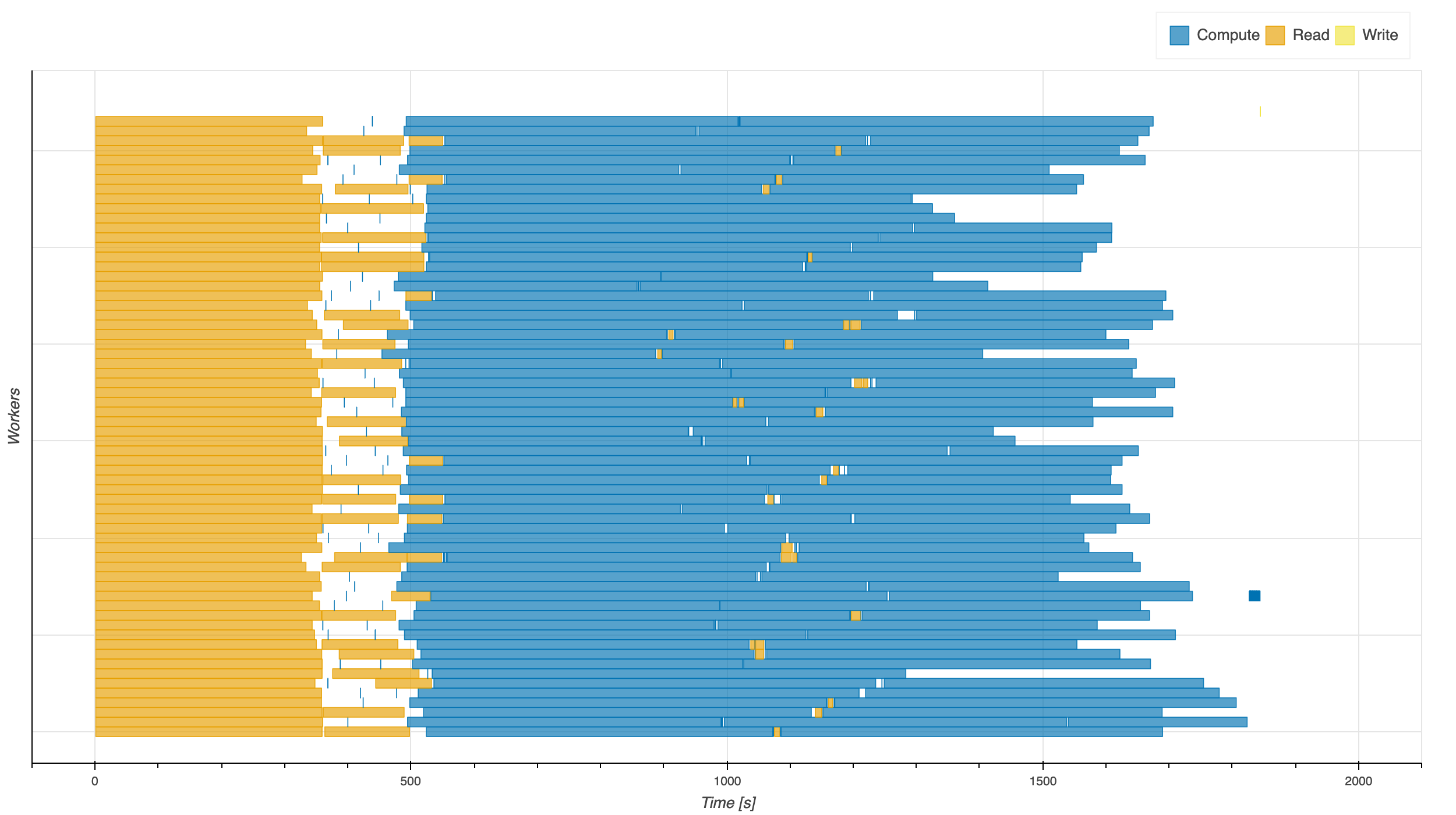}}
        \caption{Dask Delayed}\label{fig:histo_dask_delayed_gantt}
    \end{subfigure}
    \caption{Histogram (Python) Gantt chart: 125 blocks, 8 workers, 8
    CPUs/worker}\label{fig:histo_gantt}
\end{figure}

Figure~\ref{fig:histo_gantt} shows the Gantt chart of the baseline
Histogram experiment implemented in Python. Spark overhead is distributed
through all the execution while Bag overhead happens mostly when a thread
reads subsequent blocks and Delayed overhead is concentrated after reading
all the blocks.


\subsection{Histogram (NumPy): Number of workers}
\begin{figure}[!t]
    \centering
    \begin{subfigure}[b]{\columnwidth}
        \includegraphics[clip,width=\columnwidth]{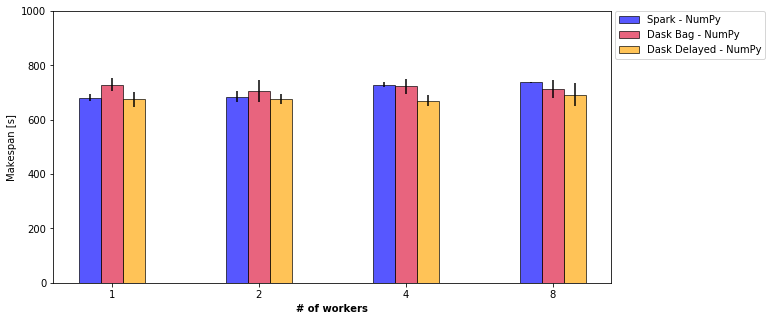}%
        \caption{Histogram (NumPy) makespan}\label{fig:histo_np_ms_worker}
    \end{subfigure}
    \\
    \begin{subfigure}[b]{\columnwidth}
        \includegraphics[clip,width=\columnwidth]{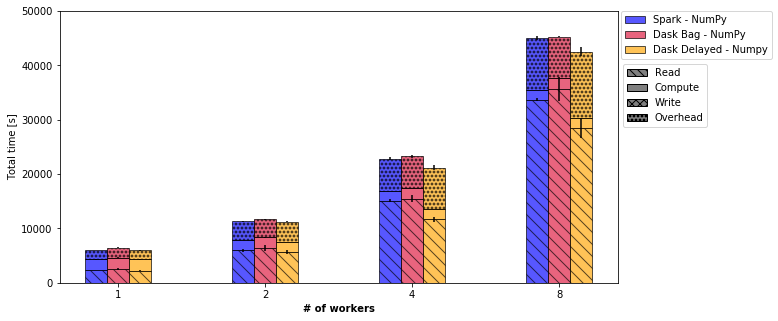}%
        \caption{Histogram (NumPy) total time}\label{fig:histo_np_tt_worker}
    \end{subfigure}
    \caption{125 blocks, 8 CPUs/worker}\label{fig:histo_np_worker}
\end{figure}

Figure~\ref{fig:histo_np_ms_worker} shows the makespan for various amount of workers.
Overall, the makespan for all engines does not change considerably with an increase
in the number of workers. This is likely a result of using a non-distributed
shared file system. There is no
substantial difference between all engines although Delayed is slightly faster.
Figure~\ref{fig:histo_np_tt_worker} shows the total execution time
breakdown. The overhead and read time increase proportionally to the number
of workers thus increasing the total time. 

\subsection{Histogram (NumPy): Number of blocks}
\begin{figure}[!t]
    \centering
    \begin{subfigure}[b]{\columnwidth}
        \includegraphics[clip,width=\columnwidth]{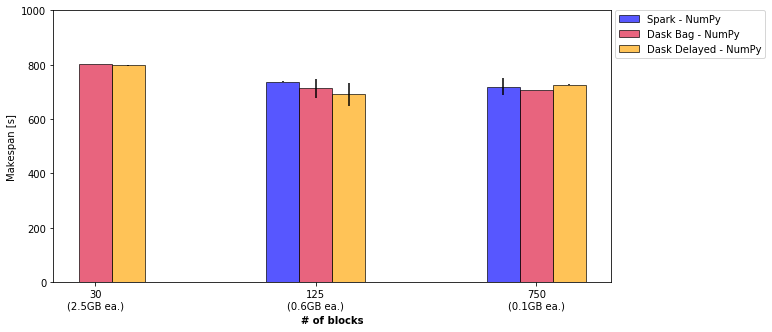}%
        \caption{Histogram (NumPy) makespan}\label{fig:histo_np_ms_block}
    \end{subfigure}
    \\
    \begin{subfigure}[b]{\columnwidth}
        \includegraphics[clip,width=\columnwidth]{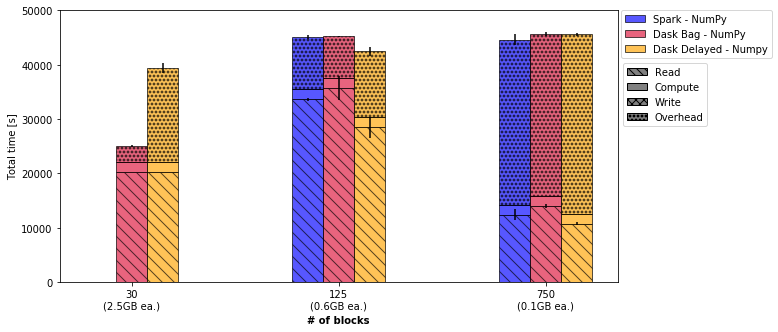}%
        \caption{Histogram (NumPy) total time}\label{fig:histo_np_tt_block}
    \end{subfigure}
    \caption{8 workers, 8 CPUs/worker}\label{fig:histo_np_block}
\end{figure}

Figure~\ref{fig:histo_np_ms_block} shows the makespan of the application for
different block sizes. Overall, there is no substantial difference for
all engines and block sizes.

Figure~\ref{fig:histo_np_tt_block} shows the total time spent in each function.
Once more, since Delayed is thread-based its overhead at 30 blocks is much
higher. Considering the overhead error and that at 30 blocks
the workers only use half their core, the total time is substantially the same
for all block size. This is due to lower block sizes having a lower data transfer time,
however, it balances out by the resulting increase in overhead.

\subsection{Histogram (NumPy): Gantt chart}
\begin{figure}[!tb]
    \centering
    \begin{subfigure}[b]{\columnwidth}
        \href{https://mathdugre.github.io/paper-big-data-engines/spark-histo_np-baseline.html}{
        \includegraphics[clip,width=\columnwidth,
        height=0.15\textheight]{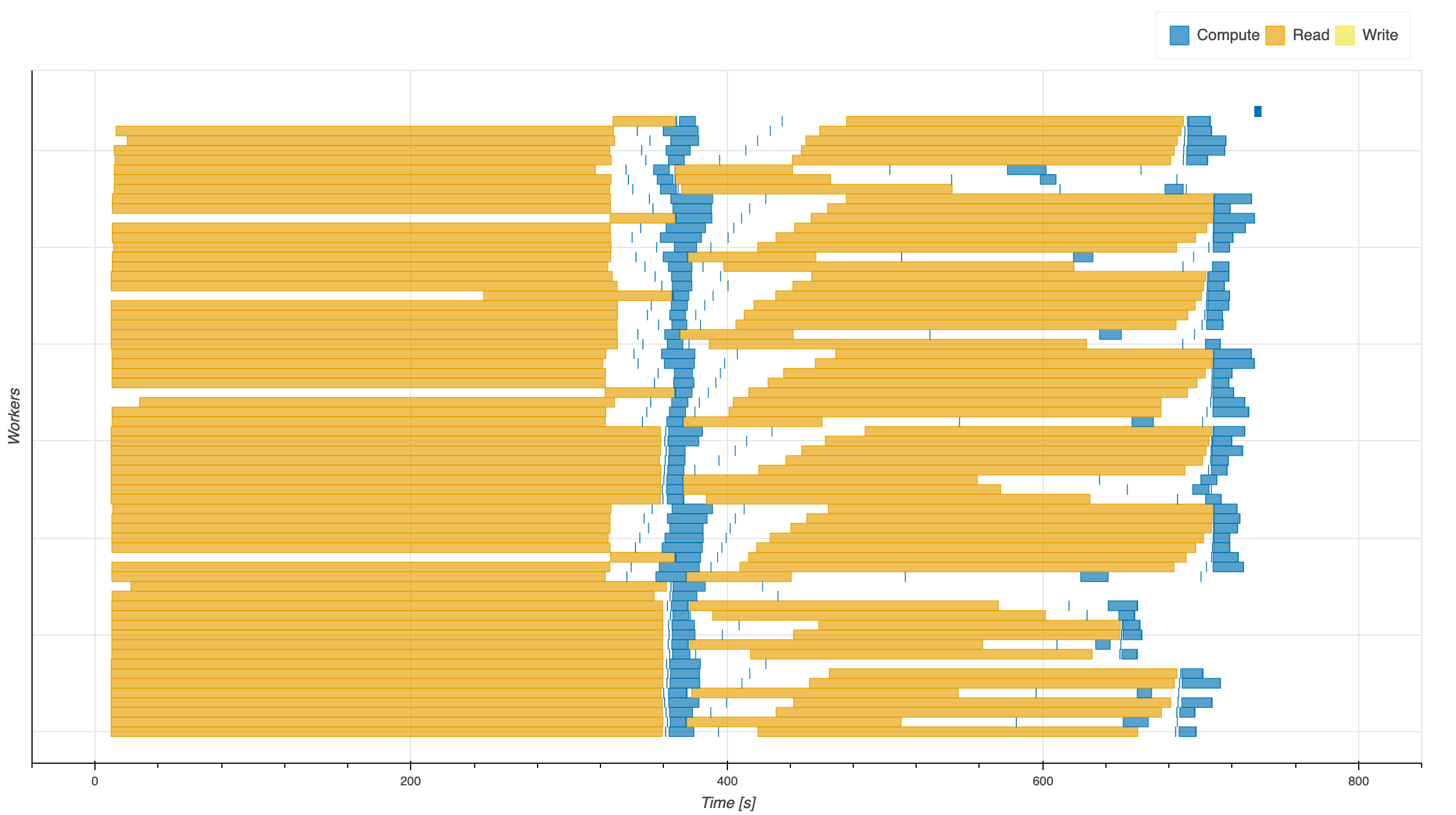}}
        \caption{Spark}\label{fig:histo_np_spark_gantt}
    \end{subfigure}
    \\
    \begin{subfigure}[b]{\columnwidth}
        \href{https://mathdugre.github.io/paper-big-data-engines/dask-bag-histo_np-baseline.html}{
        \includegraphics[clip,width=\columnwidth,
        height=0.15\textheight]{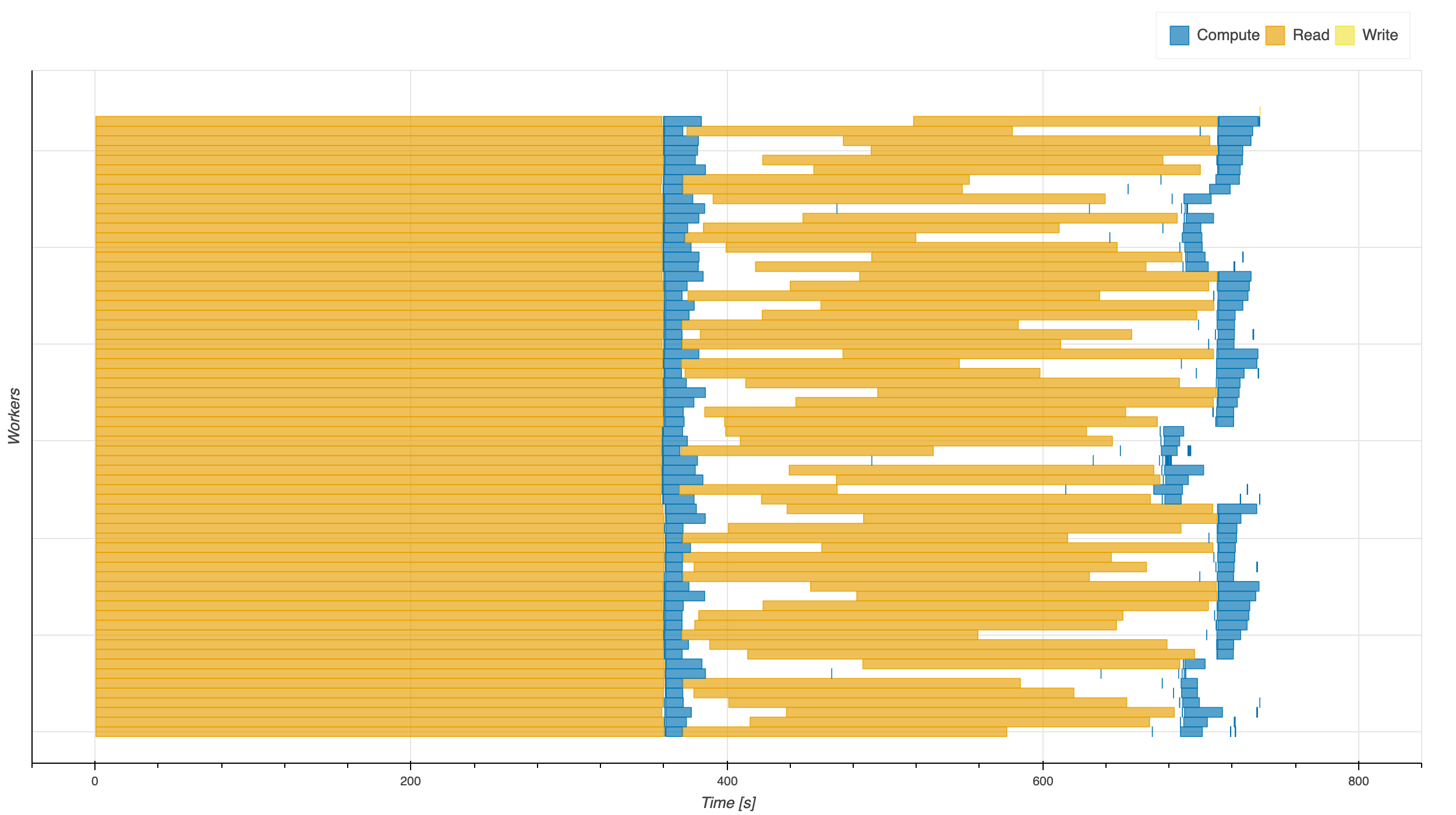}}
        \caption{Dask Bag}\label{fig:histo_np_dask_bag_gantt}
    \end{subfigure}
    \\
    \begin{subfigure}[b]{\columnwidth}
        \href{https://mathdugre.github.io/paper-big-data-engines/dask-delayed-histo_np-baseline.html}{
        \includegraphics[clip,width=\columnwidth,
        height=0.15\textheight]{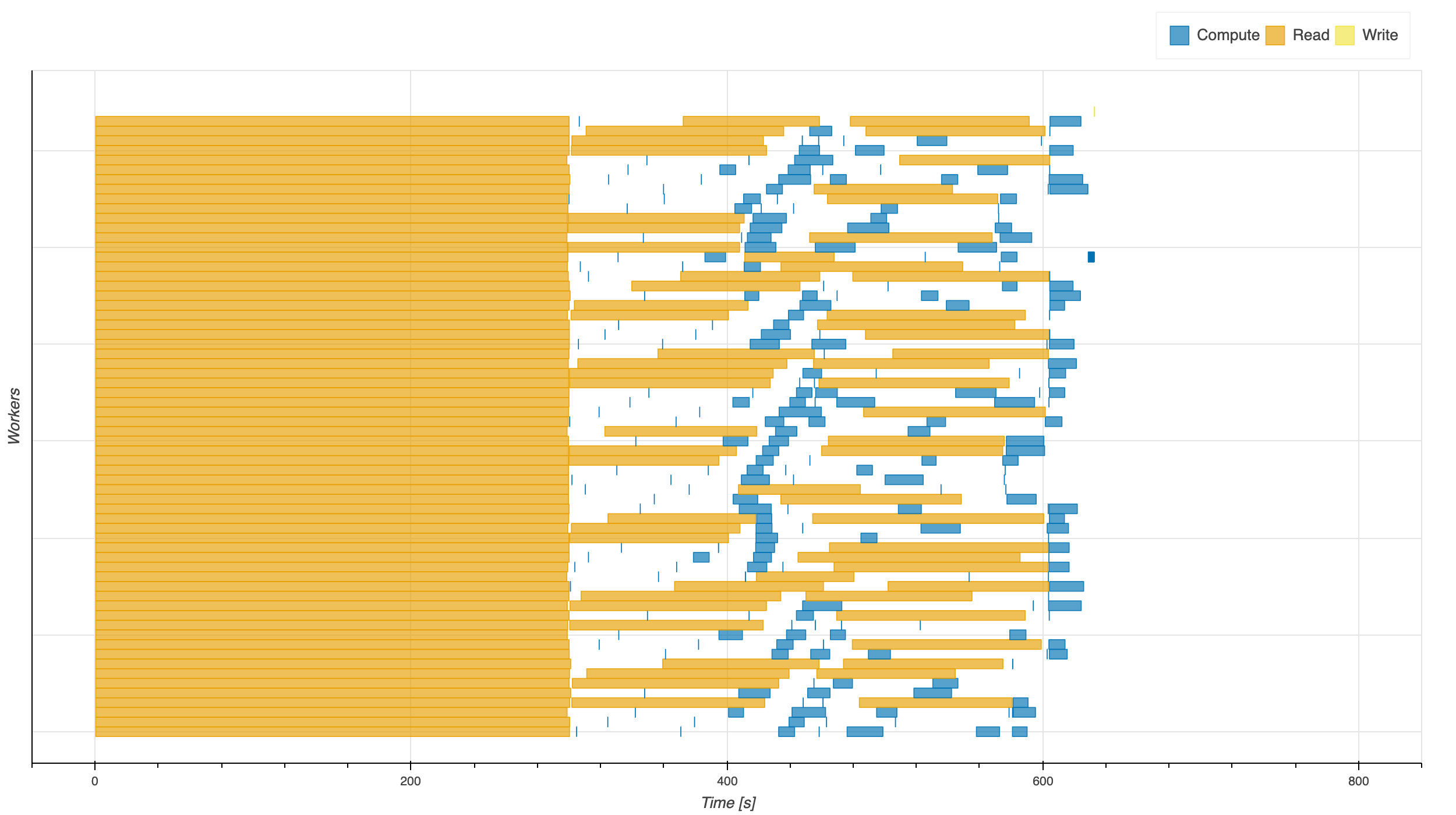}}
        \caption{Dask Delayed}\label{fig:histo_np_dask_delayed_gantt}
    \end{subfigure}
    \caption{Histogram (NumPy) Gantt chart: 125 blocks, 8 workers, 8
    CPUs/worker}\label{fig:histo_np_gantt}
\end{figure}

Figure~\ref{fig:histo_np_gantt} shows the Gantt chart of the baseline for the
Histogram (NumPy) experiment. The overhead for Spark and Bag is mostly located
between read and compute tasks. On the other hand, Delayed
overhead is dispersed between all types of tasks. This causes Delayed to
overlap more compute and read tasks simultaneously amongst different workers.

In Figure~\ref{fig:histo_np_gantt}, read tasks are initially similar for all
engines. However, when computation starts the subsequent read tasks are much
shorter for Delayed. This is due to the desynchronization of the task reducing the
data transfer bottleneck.

\subsection{BidsApp example: Number of workers}
\begin{figure}[!t]
    \centering
    \begin{subfigure}[b]{\columnwidth}
        \includegraphics[clip,width=\columnwidth]{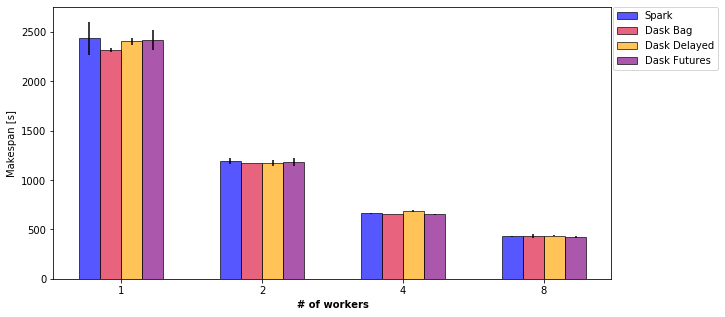}%
        \caption{BIDS App example makespan}\label{fig:bids_ms_worker}
    \end{subfigure}
    \\
    \begin{subfigure}[b]{\columnwidth}
        \includegraphics[clip,width=\columnwidth]{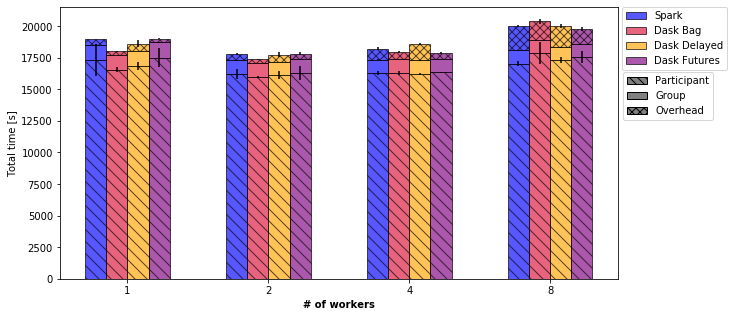}%
        \caption{BIDS App example total time}\label{fig:bids_tt_worker}
    \end{subfigure}
    \caption{Variation of the amount of worker, 8 CPUs/worker}
\end{figure}

Figure~\ref{fig:bids_ms_worker} shows the makespan of the application when
varying the number of workers. Overall, there is no substantial difference
between the engines. The makespan scales well with the increase in workers,
however, the scaling reduces slightly when reaching 8 workers. This is due
to an increase in I/O time caused by parallel accesses to 
the shared file system.

In Figure~\ref{fig:bids_tt_worker}, the total execution time of each function is
shown. Each engine has a similar total time independently of its number of workers.
This is due to the, almost, linear scaling of the engines. The overhead increases
proportionally to the number of workers with regression slopes: Spark
(\SI{224}{\second/task}), Bag (\SI{183}{\second/task}), Delayed
(\SI{173}{\second/task}), Futures (\SI{132}{\second/task}). This is due to more
communication between the scheduler and workers as well as intra-worker and
inter-worker communication.

\subsection{BidsApp example: Gantt chart}
\begin{figure}[!htb]
    \centering
    \begin{subfigure}[b]{\columnwidth}
        \href{https://mathdugre.github.io/paper-big-data-engines/spark-bids-baseline.html}{
        \includegraphics[clip,width=\columnwidth,
        height=0.15\textheight]{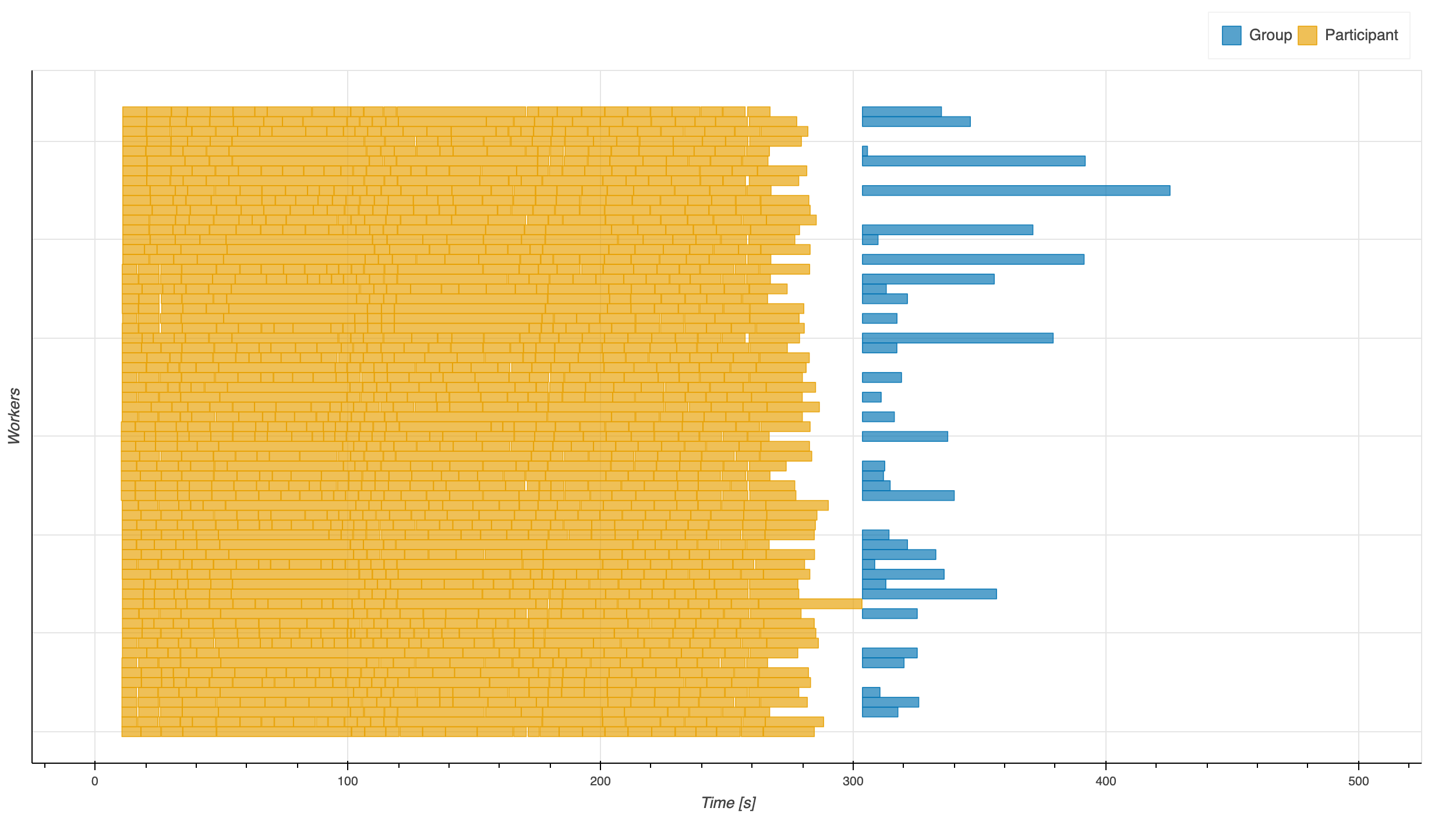}}
        \caption{Spark}\label{fig:bids_spark_gantt}
    \end{subfigure}
    \\
    \begin{subfigure}[b]{\columnwidth}
        \href{https://mathdugre.github.io/paper-big-data-engines/dask-bag-bids-baseline.html}{
        \includegraphics[clip,width=\columnwidth,
        height=0.15\textheight]{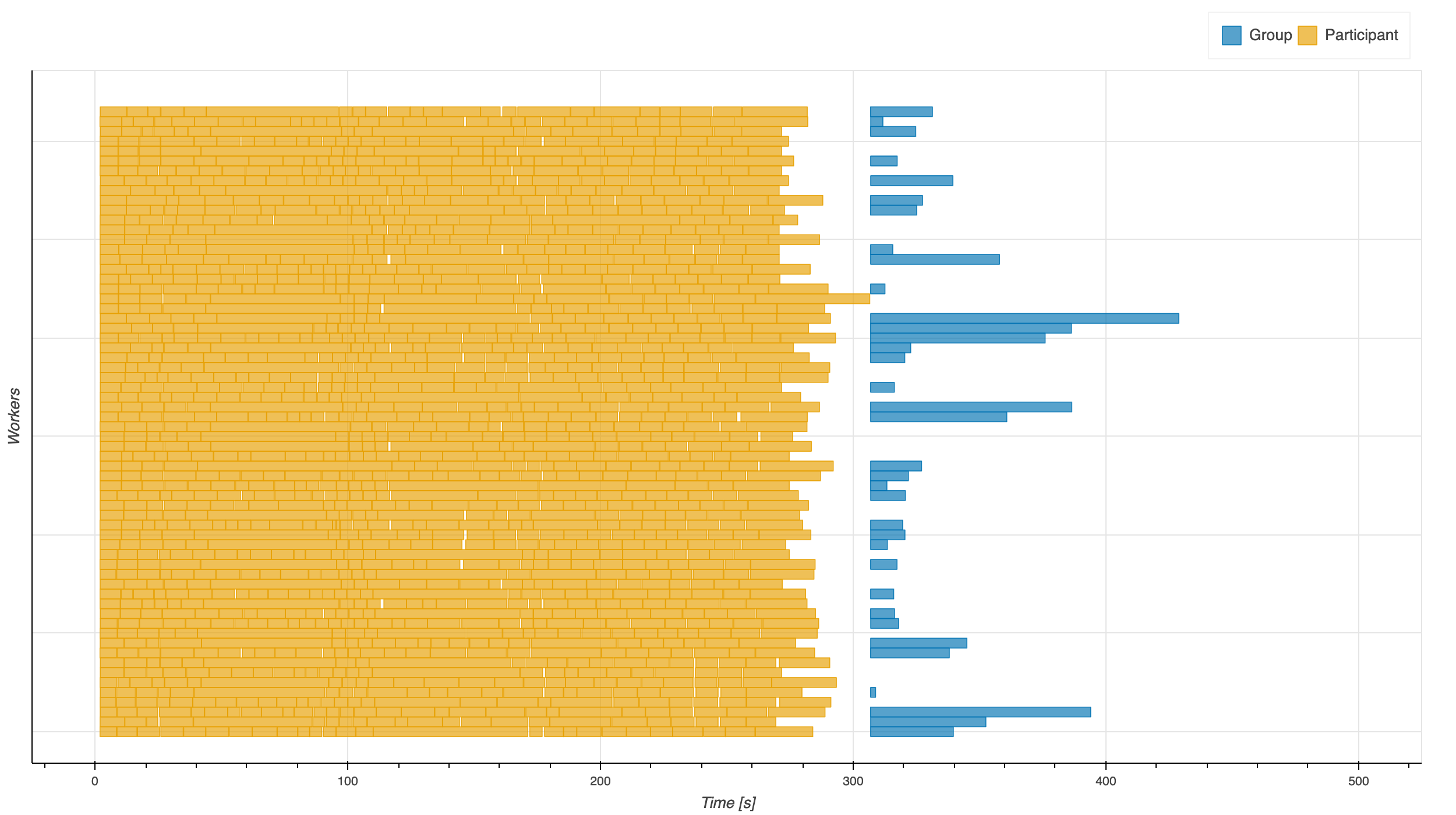}}
        \caption{Dask Bag}\label{fig:bids_dask_bag_gantt}
    \end{subfigure}
    \\
    \begin{subfigure}[b]{\columnwidth}
        \href{https://mathdugre.github.io/paper-big-data-engines/dask-delayed-bids-baseline.html}{
        \includegraphics[clip,width=\columnwidth,
        height=0.15\textheight]{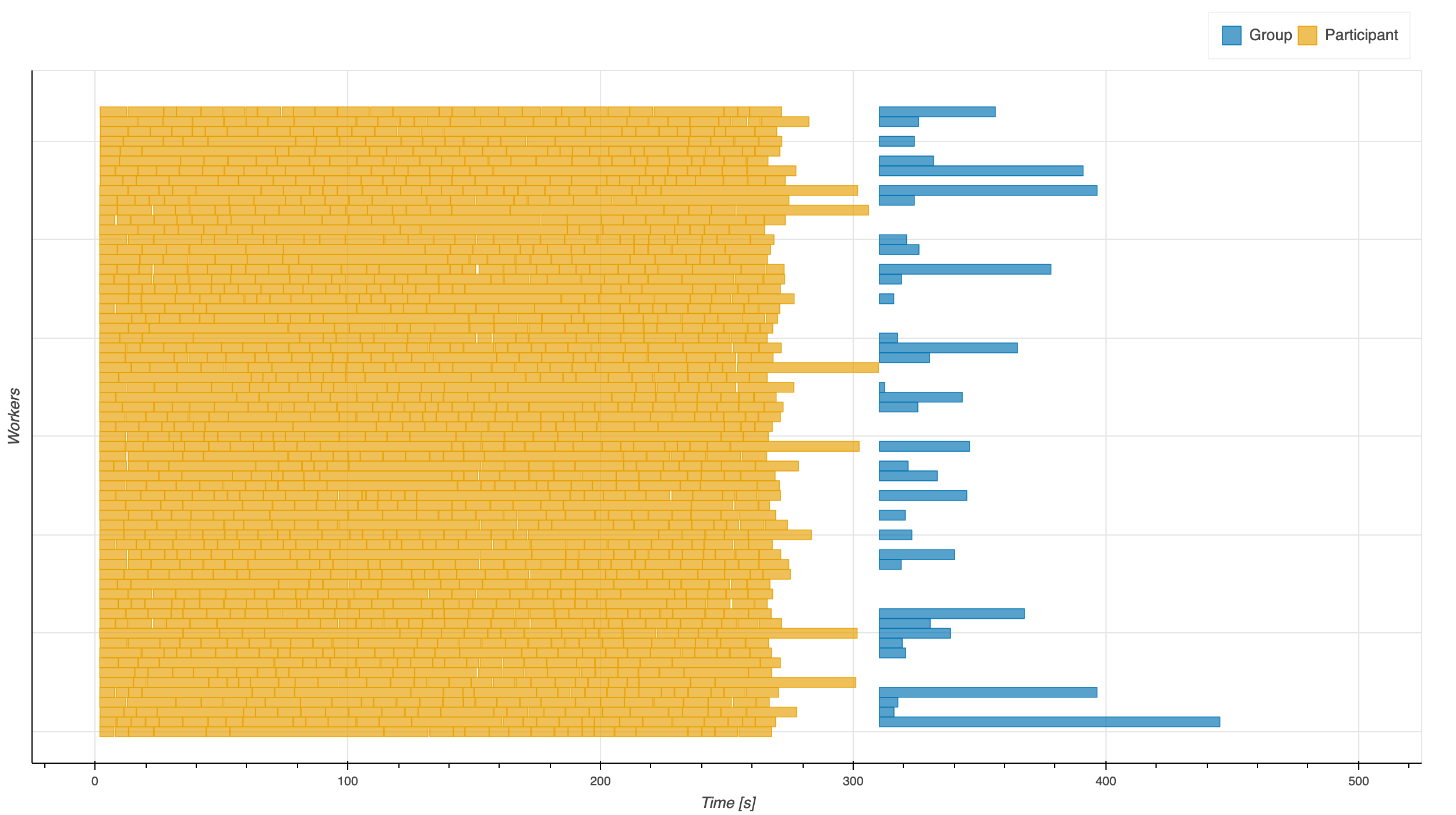}}
        \caption{Dask Delayed}\label{fig:bids_dask_delayed_gantt}
    \end{subfigure}
    \\
    \begin{subfigure}[b]{\columnwidth}
        \href{https://mathdugre.github.io/paper-big-data-engines/dask-futures-bids-baseline.html}{
        \includegraphics[clip,width=\columnwidth,
        height=0.15\textheight]{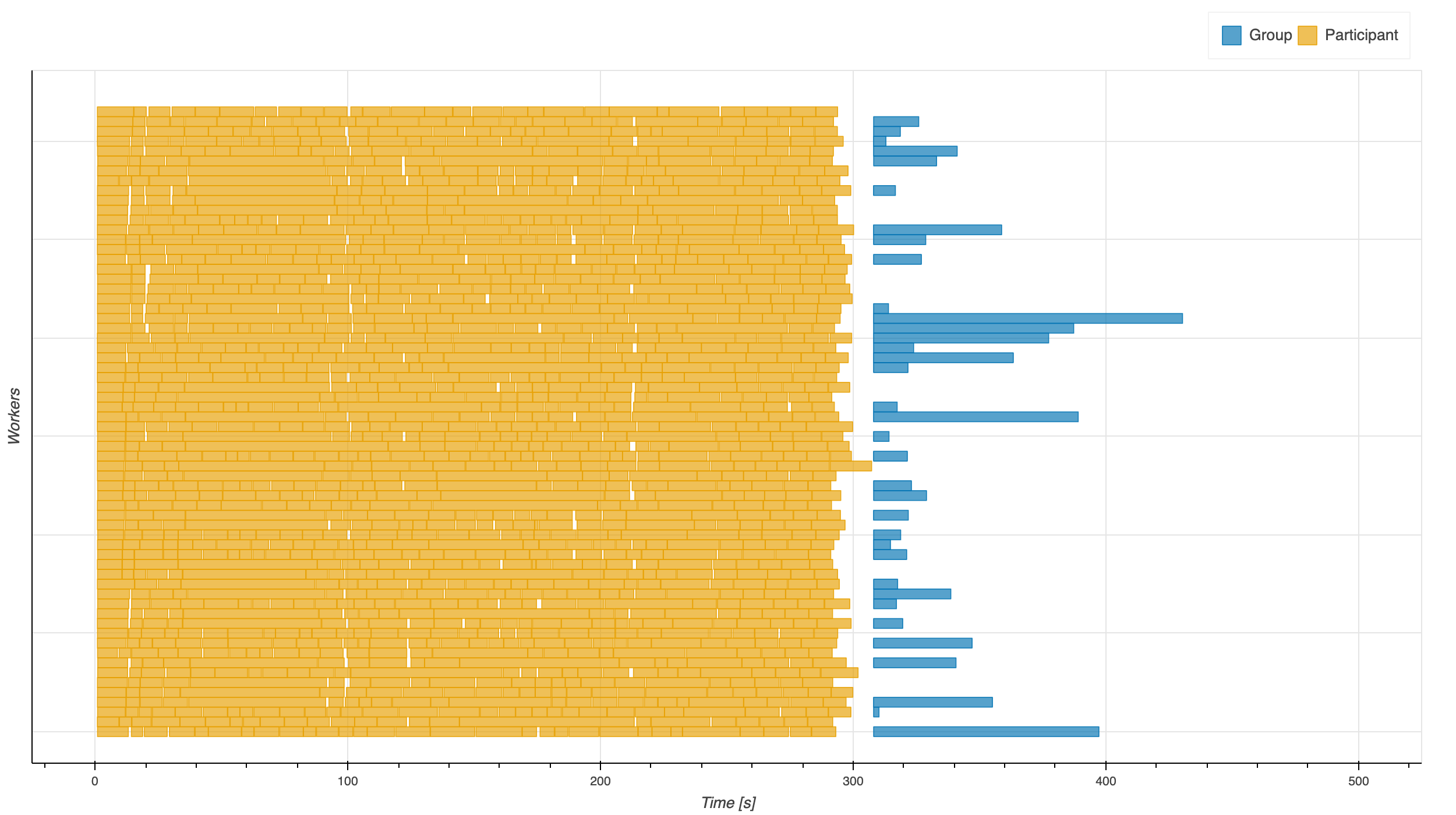}}
        \caption{Dask Futures}\label{fig:bids_dask_futures_gantt}
    \end{subfigure}
    \caption{BIDS App example Gantt chart: 8 workers}\label{fig:bids_gantt}
\end{figure}

Figure~\ref{fig:bids_gantt} shows the Gantt chart obtained for each engine.
Gantt charts are structured in two parts: participant analysis (orange) and
group analysis (blue). The participant analysis tasks differ greatly in length. This
is due to the unequal amount of sessions per subject processed. The group analysis
is similar for all engines and APIs. Overall, most of the overhead encountered
results from the transition between the two analysis. This is because the group
analysis requires the results of the participant analysis to start.

\section{Discussion}

\subsection{Comparison with previous studies}

Overall, our experiments did not show any substantial performance
difference between Spark and Dask. In contradiction with the experiments
of~\cite{Mehta:17}, we did not observe consequential overhead differences
due to serialization or startup costs. This apparent contradiction is most
likely coming from differences in applications, engine parametrization, or
infrastructure characteristics between the two studies. In our case, potential overhead differences were
compensated by reverse trends in data transfer time. Since our focus is
Big Data applications, this leads us to conclude that engines are
equivalent.

\subsection{Overhead / data transfer compensation}

The observed almost exact compensation between differences in engine
overheads and in data transfer times (see for instance
Figure~\ref{fig:inc_tt_worker}) is not surprising. Indeed, when the transfer
bandwidth is saturated, and in the absence of service thrashing,
desynchronizing file transfers usually does not reduce the makespan of data
transfers. To take an extreme example, the makespan of $n$ concurrent file
transfers of size $F$ on a system of bandwidth $B$ would be $nF/B$, as $n$
transfers would share the bandwidth, which equals to the time required
to transfer the $n$ files sequentially. In other words, saturated bandwidths
give room for extra engine overhead as long as it desynchronizes data
transfers. This could be an interesting idea to explore in data transfer optimization.

\subsection{Effect of the shared file system}

We used NFS as our shared file system as it is a common and
easy-to-configure solution to share data between compute nodes. The
performance of NFS, however, degrades substantially when accessed by
concurrent tasks, as shown in the Gantt charts in
Figures~\ref{fig:inc_gantt}, ~\ref{fig:histo_gantt}, ~\ref{fig:histo_np_gantt} 
and~\ref{fig:bids_gantt}. This is due to the limitations imposed by network and disk
bandwidths. While network bandwidth can hardly be overcome, disk bandwidth
can be increased by storing data to multiple disks addressed by a parallel
file system such a Lustre~\cite{lustre}. Such a configuration would likely
reduce the impact of data transfers, although the main observed patterns
should remain.

Throughout the applications and engines, the use of large data blocks amplified the
consequences of using an NFS. Moreover, the bandwidth saturation almost neglects the
scaling, from additional workers, for data-intensive applications: Incrementation and
Histogram (NumPy). On the other hand, an almost linear scaling shows that it has no
substantial effect on compute-intensive applications: Histogram (Python) and BIDS app
example.

We suppose that using a parallel file system would allow applications to have almost
linear scaling as long as they do not saturate the bandwidth of the system.

\subsection{Effect of Python's Global Interpreter Lock}

Although Dask reduces the impact of Python's GIL on parallelism by using external
Python extensions, our Python histogram experiment (Figure
~\ref{fig:histo_worker} and \ref{fig:histo_block}) showed that issues remain in Dask to
parallelize Python functions. However, Dask was easily able to parallelize
the same application when implemented with NumPy
(Figure~\ref{fig:histo_np_worker} and \ref{fig:histo_np_block}) because in this case the computation
happened in C, outside of the GIL. Similarly, the computation in
command-line applications happens in a sub-process, hence outside of the
GIL, which explains why the BIDS apps example application was not impacted
by this issue. Although implemented in Python, Incrementation was not
impacted either, as computation was only emulated through sleep time.

This behavior is also the consequence of our configuration of Dask that
created 1 worker process with 8 threads. Since the Histogram Python
implementation does not release the GIL it can only use one of the thread
at a given time thus it slows down the computation significantly. 
Our results suggest configuring Dask with multi-threaded workers for application
releasing the GIL and multi-process workers while the default configuration of Spark
workers brings stable results across different types of applications.

\subsection{Effect of data partitioning}

The number of data partitions also has an impact on application
performance. In our experiments, we set a partition for each block or
subject, resulting in a maximum of 750 partitions for the BigBrain
applications, and 1,397 partitions for the BIDS app. The number of
partitions obeys a classical trade-off: small numbers of partitions reduce 
overhead, while large numbers of
partitions increase parallelism. Since both Spark and Dask can
work with custom partition sizes, we do not expect substantial differences
coming from variations of this parameter.


\section{Conclusion}
We presented a comparison of two Big Data engines, Apache Spark and Dask.
We studied the engines on three data-intensive neuroimaging applications
representative of common use cases. Overall, our results show no substantial
performance difference between the engines. Interestingly, differences in
engine overheads do not impact performance due to their impact on data
transfer time: higher overheads are almost exactly compensated by a lower
transfer time when data transfers saturate the bandwidth. These results
suggest that future research should focus on strategies to reduce the
impact of data transfers on applications.

\section*{Acknowledgment}

Mathieu Dugr\'e was funded by an Undergraduate Student Research Assistant award from
the National Science and Engineering Research Council of Canada. We warmly thank
Compute Canada and its regional center Westgrid for providing the cloud infrastructure used in
these experiments, and the McGill Center for Integrative Neuroscience for giving us
access to their cloud allocation.

\bibliographystyle{IEEEtran}
\bibliography{IEEEabrv,reference}

\end{document}